%% file: root.tex
\documentclass{article}

\PassOptionsToPackage{numbers, compress}{natbib}
\usepackage[preprint]{neurips_2025}

\usepackage[utf8]{inputenc} 
\usepackage[T1]{fontenc}    
\usepackage{hyperref}       
\usepackage{url}            
\usepackage{booktabs}       
\usepackage{amsfonts}       
\usepackage{nicefrac}       
\usepackage{microtype}      
\usepackage{xcolor}         
\usepackage{graphicx}
\usepackage{subfigure}
\usepackage{booktabs}
\usepackage{multirow}
\usepackage{multicol}
\usepackage{comment}
\usepackage{tablefootnote}
\usepackage{framed}
\usepackage{diagbox}
\usepackage{algorithm}
\usepackage{algorithmic}
\usepackage{subcaption}
\usepackage{arydshln}

\usepackage{hyperref}

\usepackage{amsmath}
\usepackage{amssymb}
\usepackage{mathtools}
\usepackage{amsthm}

\usepackage[capitalize,noabbrev]{cleveref}

\usepackage[textsize=tiny]{todonotes}

\theoremstyle{plain}

\theoremstyle{definition}

\theoremstyle{remark}

\newtoggle{showcomments}
\togglefalse{showcomments}

\newcommand{\definecommenter}[2]{%
  \newcommand{#1}[1]{%
    \iftoggle{showcomments}{%
      \textcolor{#2}{{{\detokenize{#1}}}: \textit{##1}}%
    }{}%
  }%
}

\definecommenter{\yubeen}{cyan}
\definecommenter{\minchan}{orange}
\definecommenter{\jaejin}{purple}
\definecommenter{\sangbum}{yellow}
\definecommenter{\jaehyung}{green}
\definecommenter{\yejin}{red}
\definecommenter{\niloofar}{teal}

\title{
Privacy-Preserving LLM Interaction with \\
Socratic Chain-of-Thought Reasoning and \\
Homomorphically Encrypted Vector Databases
}

\author{
  \textbf{
  Yubeen Bae$^{1}{\thanks{
  Equal contribution (alphabetical order). Code is available at \href{https://github.com/Yubeen-Bae/PPMI}{\texttt{https://github.com/Yubeen-Bae/PPMI}}.
  }}$ \quad
  Minchan Kim$^{1*}$ \quad
  Jaejin Lee$^{1*}$ \quad
  Sangbum Kim$^{1}$ \quad
  Jaehyung Kim$^{2}$}
  \\[0.21em]
  \textbf{
  Yejin Choi$^{3}$ \quad
  Niloofar Mireshghallah$^{4}$}
  \\[0.21em]
  $^{1}$Seoul National University \quad
  $^{2}$Stanford University \quad
  $^{3}$NVIDIA \quad
  $^{4}$University of Washington
  \\[0.21em]
  \texttt{\{lights0320, kjkk0502, jaejin.lee\}@snu.ac.kr}
  \vspace{-1.6em}
}

\begin{document}

\maketitle


\input{sections/0_abstract}
\input{sections/1_introduction}
\input{sections/2_background}
\input{sections/3_framework}
\input{sections/4_database}
\input{sections/5_experiment}
\input{sections/6_related_work}
\input{sections/7_conclusion}
\input{sections/8_acknowledgments}


{
\small
\bibliographystyle{plainnat}
\bibliography{refs}
}


\newpage
\appendix
\input{sections/A_database}
\input{sections/B_experimental_setup}
\input{sections/C_compute_resources}
\input{sections/D_qualitative_analysis}
\input{sections/E_prompt_templates}



\end{document}

%% file: sections/0_abstract.tex
{\centering \input{figures/framework} \par}
\begin{abstract}
Large language models (LLMs) are increasingly used as personal agents, accessing sensitive user data such as calendars, emails, and medical records.
Users currently face a trade-off: They can send private records—many of which are stored in remote databases—to powerful but untrusted LLM providers, increasing their exposure risk. Alternatively, they can run less powerful models locally on trusted devices.
We bridge this gap: Our \textbf{Socratic Chain-of-Thought Reasoning} first sends a generic, non-private user query to a powerful, untrusted LLM, which generates a Chain-of-Thought (CoT) prompt and detailed sub-queries without accessing user data. Next, we embed these sub-queries and perform encrypted sub-second semantic search using our \textbf{Homomorphically Encrypted Vector Database} across one million entries of a single user's private data. This represents a realistic scale of personal documents, emails, and records accumulated over years of digital activity.
Finally, we feed the CoT prompt and the decrypted records to a local language model and generate the final response.
On the LoCoMo long-context QA benchmark, our \textbf{hybrid framework}—combining GPT-4o with a local Llama-3.2-1B model—outperforms using GPT-4o alone by up to 7.1 percentage points. This demonstrates a first step toward systems where tasks are decomposed and split between untrusted strong LLMs and weak local ones, preserving user privacy.
\end{abstract}

%% file: figures/framework.tex
\begin{figure}[h]
\centering
\includegraphics[width=\textwidth]
{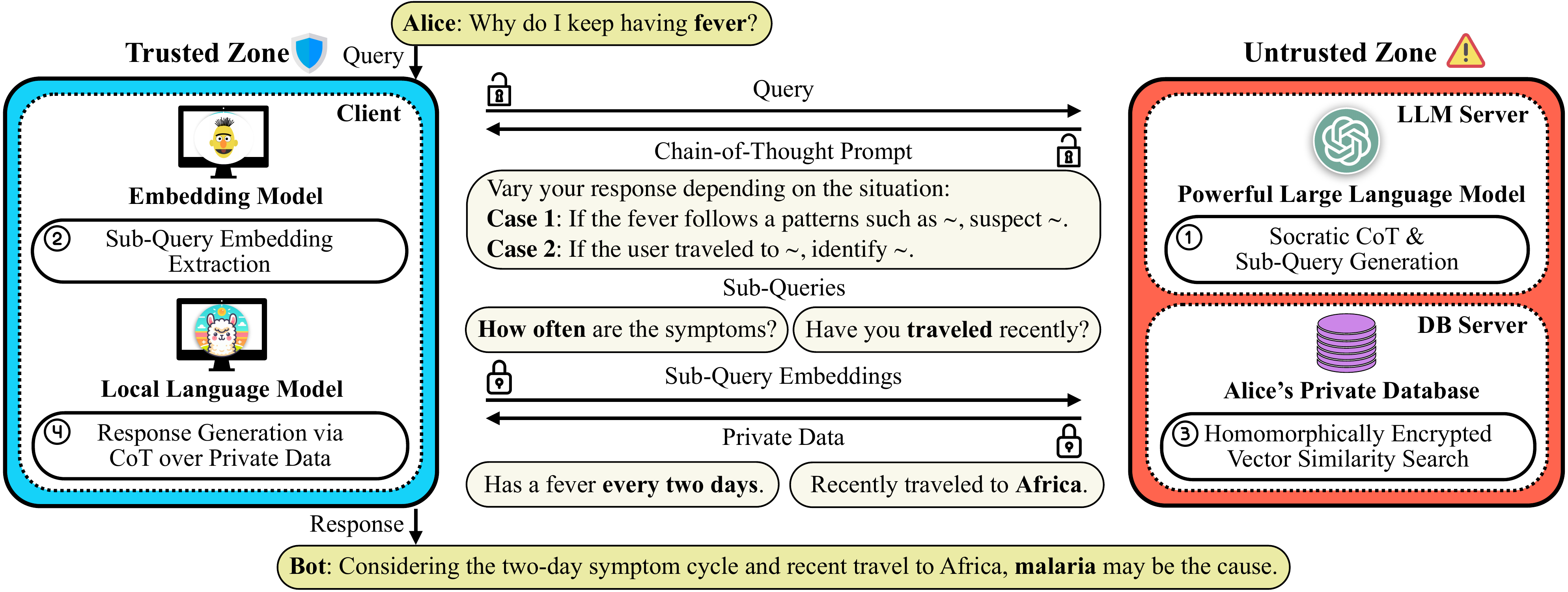}
\caption{
\textbf{Overview of our hybrid framework.} Upon receiving a query, a remote LLM generates a Chain-of-Thought (CoT) prompt and sub-queries (Stage~1) which are embedded locally (Stage~2), and used for our encrypted vector search on a remote database (Stage~3).
Retrieved records are decrypted and provided with the CoT prompt as context to a local model to generate the final response (Stage~4).
}
\label{fig:overview}
\end{figure}

%% file: sections/1_introduction.tex
\section{Introduction}
\label{sec:introduction}

Large language models (LLMs) are becoming the default backend for personal agents that manage emails, schedule meetings, and process real-time health data from wearable devices~\cite{qiu2024llm,liuagentbench,song2025togedule}. These agents must integrate data from heterogeneous sources—many stored remotely in cloud databases—using retrieval-augmented generation (RAG)~\cite{lewis2020retrieval}. While forwarding user queries along with retrieved data to powerful yet untrusted LLMs enhances performance, it introduces substantial privacy risks by potentially exposing private records~\cite{zeng2024privacy,jiang2024ragthief}. Conversely, restricting these operations to local trusted devices significantly degrades performance~\cite{liu2025edge}.
This raises the question:~\emph{Can we perform LLM interactions on private data while maintaining efficiency and accuracy without privacy risks or significant performance degradation?}

Existing privacy-preserving methods, such as data minimization or scrubbing personally identifiable information (PII), often sacrifice data utility or provide limited privacy through superficial suppressions~\cite{xin2025false}.
To bridge the privacy-utility gap, we propose a four-stage hybrid framework that clearly delineates trusted and untrusted environments (left and right sides of Figure~\ref{fig:overview}), ensuring private raw data either remains strictly within local boundaries or is securely encrypted when externally stored or searched.
We integrate two novel components: (1) \textbf{Socratic Chain-of-Thought Reasoning}, which enables challenging yet non-private queries to be offloaded to a powerful external language model; and (2) a \textbf{Homomorphically Encrypted Vector Database}, a cryptographic system that allows efficient semantic search over encrypted records without ever decrypting them. 
This enables users to leverage cloud storage and compute resources while maintaining complete privacy—the cloud provider can execute searches without learning anything about the data content or search queries.

\textbf{Stage~1}: When the user, Alice, poses a query (see Figure~\ref{fig:overview}, top), our Socratic Chain-of-Thought Reasoning elicits a detailed Chain-of-Thought prompt and sub-queries from a powerful external LLM. We provide only the main query, which we assume to be non-private in our protocol, to the external LLM without exposing any private user data. Rather than directly providing a diagnosis, we prompt the powerful LLM to generate Chain-of-Thought prompt for reasoning and targeted sub-queries for retrieval—in this case, questions about medications and travel history. This approach allows the powerful model to break down complex task into simpler ones, making it easier for the weaker local model to reason effectively when given access to private data as context. \textbf{Stage~2}: These sub-queries are then locally embedded to prepare them for secure semantic search over our encrypted vector database containing Alice's relevant records.

\textbf{Stage~3}: Once the sub-query embeddings reach our Homomorphically Encrypted Vector Database, the system executes secure vector similarity search, where all key vectors are homomorphically encrypted and compared against a million encrypted key vectors. 
Our novel inner product protocol computes similarity entirely in the encrypted domain in under one second using standard CPUs. The system then retrieves the corresponding encrypted records, returning top-\( k \) matches from a million-entry store in encrypted format.
\textbf{Stage~4}: Finally, a much smaller, weaker language model operating exclusively within the local trusted zone generates the final response, drawing on both the chain-of-thought prompt and the decrypted private records supplied by the stronger remote model.

We extensively evaluate our framework on two long-context QA benchmarks. \emph{LoCoMo} assesses recall of extensive conversational histories~\cite{locomo}, while \emph{MediQ} tests interactive clinical reasoning~\cite{li2024mediq}. We establish two baselines representing privacy extremes: (1) a local-only (fully private) baseline using Llama-3 with 1B and 3B parameters, and (2) a remote-only (fully non-private) golden baseline using GPT-4o and Gemini-1.5-Pro. Our approach provides a balanced trade-off between these extremes.

Through our \textbf{Socratic Chain-of-Thought Reasoning}, the Llama 1B-parameter local model achieves an F1 score of 87.7 on LoCoMo, notably surpassing GPT-4o by 7.1 percentage points and the local-only baseline by 23.1 percentage points. This improvement likely stems from additional test-time computation enabled by the chain-of-thought process~\cite{chen2023frugalgpt}. For MediQ, improvements are relatively smaller due to domain-specific adaptation challenges.
Our \textbf{Homomorphically Encrypted Vector Database} efficiently searches entries from $10^6$ records in under one second on commodity CPUs, maintaining > 99\% Recall@5 with a median storage overhead of just \( \mathbf{5.8\times} \). \emph{Collectively, our findings mark an important step toward privacy-preserving systems that effectively partition tasks between untrusted high-capacity LLMs and trusted lightweight local models, without requiring any additional post-training.}
\clearpage

%% file: sections/2_background.tex
\section{Background and Problem Formulation}
\label{sec:background}

Large language models (LLMs) increasingly serve as personal assistants, processing sensitive user data such as calendars, emails, and medical records~\cite{zeng2024good, qiu2024llm}. 
Effective LLM-based personal assistants require two fundamental capabilities:

\noindent\textbf{(1) Contextual Reasoning:} The model must establish clear criteria to accurately interpret user queries in context. For instance, recognizing \textit{a cyclic fever pattern recurring every two days} in combination with \textit{recent travel to Africa} strongly suggests \textit{malaria}. Augmenting such contextual understanding into the reasoning process ensures precise and meaningful conclusions.

\noindent\textbf{(2) Contextual Data Retrieval:} The model must determine which contextual data is necessary for comprehensive understanding. As illustrated in Figure~\ref{fig:overview}, a user's query such as \textit{"Why do I keep having fever?"} might not provide enough context to retrieve all necessary records. The model must generate targeted sub-queries to collect comprehensive information, such as travel history that might reveal malaria risk factors~\cite{lewis2020retrieval}.

\textbf{Privacy Problem Formulation:}
While powerful cloud-based LLMs offer superior reasoning capabilities, they require users to expose private data to untrusted providers~\cite{confaide}. Conversely, local models that preserve privacy lack the computational capacity for complex reasoning tasks. We consider a user with a non-private query whose answer depends on private records stored remotely (As shown in Figure~\ref{fig:overview}). The local device has limited computational resources insufficient for complex reasoning, while powerful cloud LLMs cannot be trusted with sensitive data~\cite{wang2024cloud}.

\textbf{Threat Model:} We protect against three adversaries: (1) the LLM provider who receives user queries, (2) the database provider storing encrypted records~\cite{bonnetain2019quantum}, and (3) external attackers who may compromise these services~\cite{hutchins2011intelligence}. Even with standard encryption, providers typically hold decryption keys, enabling potential privacy breaches through insider threats or security compromises~\cite{cappelli2012cert, hunker2008insiders}.

\textbf{Privacy Goal:} User data must remain encrypted outside the trusted local environment, with decryption keys never leaving the user's control. The system must enable complex reasoning and efficient retrieval while ensuring that untrusted components cannot access plaintext private data~\cite{gentry2009fully, rivest1978data}.

%% file: sections/3_framework.tex
\section{Privacy-Preserving Framework with Socratic Chain-of-Thought Reasoning}
\label{sec:socratic}

We present our framework that enables powerful LLM reasoning while maintaining strict privacy guarantees, ensuring that sensitive user data is never exposed during interaction.
This section describes our overall approach, which combines Socratic Chain-of-Thought Reasoning with our privacy-preserving framework designed to separate trusted and untrusted zones (Section~\ref{sec:framework-overview}).
Section~\ref{sec:database} further details our homomorphically encrypted retrieval system, which supports secure access to private records without compromising confidentiality.

\subsection{Framework Overview}
\label{sec:framework-overview}

Figure~\ref{fig:overview} illustrates our framework's architecture, which separates computation into trusted and untrusted zones to balance privacy and performance. In the trusted zone (left side of the figure), the user's local device hosts a lightweight language model and embedding model with exclusive access to decryption keys, ensuring that sensitive data never leaves the user's control in plain form. The untrusted zone (right side of the figure) comprises cloud providers hosting: (1) a powerful LLM for abstract reasoning, and (2) an encrypted vector database storing the user's private records using homomorphic encryption~\cite{gentry2009fully,brakerski2014leveled}, allowing secure processing without data decryption.

Consider the medical consultation example in Figure~\ref{fig:overview}: when a user asks ``Why do I keep having fever?'', the query flows to the remote LLM without exposing any private medical history. The powerful model generates targeted sub-queries (e.g., symptom frequency, travel history) that guide retrieval from the encrypted database, where personal records remain protected even during search operations thanks to homomorphic encryption. This architectural separation provides both active control—users explicitly manage what reaches remote models—and passive control, where cryptographic protection ensures data remains secure even if users make mistakes~\cite{goldreich2019foundations}.

\subsection{Framework Operation}
\label{sec:framework-operation}
The following detailed example illustrates our framework's operation, as shown in Figure~\ref{fig:overview}:
\begin{enumerate}
    \item The process begins when a user submits a query $x$: 
    \begin{center}
    \textit{Why do I keep having fever?}
    \end{center}
    
    \item Given the user's input $x$, the remote LLM generates:
    \begin{itemize}
        \item A Chain-of-Thought (CoT) prompt $c$ via its CoT generator $G_{c}$: 
        \begin{center}
        \begin{tabular}{l}
        \textit{Vary your response depending on the situation:}\\
        \textit{Case 1: If the fever follows patterns such as \textasciitilde, suspect \textasciitilde.}\\
        \textit{Case 2: If the user traveled to \textasciitilde, identify \textasciitilde.}
        \end{tabular}
        \end{center}
        
        \item Relevant sub-queries via its sub-query generator $G_{q}$:
        \begin{center}
        \textit{How often are the symptoms?}\\
        \textit{Have you traveled recently?}
        \end{center}
    \end{itemize}
    
    \item The local client embeds these sub-queries and executes encrypted search on the user's private database $\mathcal{D}$, using a retriever $R$ to obtain records $v$:
    \begin{center}
    \textit{Has a fever every two days.}\\
    \textit{Recently traveled to Africa.}
    \end{center}
    
    \item Finally, the local model $L$ integrates the CoT prompt $c$ and retrieved records $v$ to generate the final response $y$:
    \begin{center}
    \textit{Considering the two-day symptom cycle and recent travel to Africa,\\ malaria may be the cause.}
    \end{center} 
\end{enumerate}

We have provided examples of our prompts and the chains in Appendix~\ref{sec:qualitative_analysis}. To formalize this process, let $\mathcal{V}$ be the set of tokens and define $k$-tuples of $\mathcal{V}$ as:
\[
\mathcal{V}^{k} = \{ (v_{0}, \ldots, v_{k-1}) \mid v_{0}, \ldots, v_{k-1} \in \mathcal{V} \}
\]
Then $\mathcal{V}^{*} = \bigcup_{k=0}^{\infty} \mathcal{V}^{k}$ is the set of all finite-length sequences.

We denote the Chain-of-Thought generator as $G_{c}$, the sub-query generator as $G_{q}$, and the retriever as $R$, which operates on a database $\mathcal{D}$. The local model $L$ generates a response $y$ based on an input $x$, using the CoT prompt and retrieved records as context:
\[
y = L(x, c, v, h)
\]
where $x, y, c, v, h \in \mathcal{V}^{*}$. Specifically:
\begin{itemize}
\item $x$ denotes the user's input query
\item $c = G_{c}(x)$ represents the CoT prompt generated by the remote model
\item $v = R(G_{q}(x), \mathcal{D})$ indicates the retrieved records obtained by querying the encrypted database $\mathcal{D}$ with sub-queries generated by $G_{q}$
\item $h$ represents optional historical context (e.g., previous conversation turns), defaulting to an empty tuple $()$ if not provided
\item $y$ is the final response generated by the local model
\end{itemize}

This decomposition ensures that remote models $G_c$ and $G_q$ operate only on non-private data, while private records in $\mathcal{D}$ remain encrypted and are processed only within the trusted local environment by $L$. The next section details our homomorphically encrypted vector database that enables efficient retrieval without compromising privacy.

%% file: sections/4_database.tex
\section{Homomorphically Encrypted Vector Database}
\label{sec:database}

In this section, we discuss the design and implementation of a vector database operating over encrypted data, integrating Homomorphic Encryption (HE) and Private Information Retrieval (PIR) techniques to enable secure and efficient semantic search with rapid updates. We begin by exploring the necessity for remote encrypted vector databases and their setup. Subsequently, we analyze existing HE-based inner product (IP) computations, proposing enhancements that significantly improve efficiency with faster updates. Finally, we present a detailed framework and its corresponding API specifications, illustrated clearly through algorithmic tables.

\subsection{Motivations and Setup}

The performance of personal assistants powered by language models significantly improves when relevant user-context data is appropriately provided. Thus, seamless integration and accumulation of user data are crucial for developing powerful personal assistants. While storing data locally on the user’s device allows quick retrieval, local storage is inherently limited in capacity. Consequently, leveraging cloud-based solutions becomes essential, offering extensive storage capabilities and seamless data accumulation. Moreover, cloud solutions effectively handle multi-device scenarios by integrating data from diverse sources, such as wearable devices, and provide a unified environment, simplifying overall data management compared to fragmented local storage approaches.

Consider a simple yet inefficient baseline protocol for implementing a remote encrypted vector database: whenever a client needs to search or update an entry, it downloads the entire database from the server, decrypts it locally, performs the necessary operations, encrypts the entire database again, and uploads it back to the server. Although straightforward, this approach incurs significant communication overhead and computational burden on the client, rendering it impractical for large-scale applications.

To address these inefficiencies, we aim to design a remote vector database system that maintains the same robust security guarantees as the naive approach—where the database remains encrypted under a symmetric key held exclusively by the client—but achieves significantly better efficiency in communication and client-side computation.

The retrieval process within a vector database typically involves two critical sub-processes: search and return. The search phase computes similarity scores between a query vector and the key vectors stored in the database, and selects the top-\( k \) most relevant entries. In the return phase, corresponding data values are fetched from the database based on the selected entry identifiers (ids).

To ensure the robust security level of the baseline, three main processes must be executed in an oblivious manner: inner product (IP) computations, top-\( k \) selection, and data access. Homomorphic encryption (HE) is particularly effective for inner product calculations, as it significantly reduces both communication rounds and client-side computation. However, top-\( k \) selection, involving numerous logical comparisons, becomes computationally intensive when directly implemented with HE~\cite{kway-sorting}. Therefore, we adopt a client-aided approach, enabling the client to efficiently select the top-\( k \) entry ids without excessive computational overhead.

Finally, once the client identifies the relevant entry ids, records are securely retrieved under the same security guarantees as the naive baseline.
We use private information retrieval (PIR) protocols to fetch values corresponding to these ids without revealing which database entries are accessed.

To further enhance the efficiency of the database, particularly regarding frequent updates, we propose a novel HE-based IP algorithm that balances rapid updates and efficient search performance. Existing sublinear PIR schemes that rely on preprocessing are impractical for dynamic databases due to high preprocessing costs~\cite{hintlesspir}. To mitigate this issue, there are several line of works~\cite{spiralpir, onionpir}, single-server PIR protocol that operates efficiently without preprocessing, thereby enabling dynamic and rapid updates alongside secure and efficient searching.

\subsection{Secure Inner Product, Technical Overview}
\label{subsec:encrypted_vector_db_sip}

Given a power-of-two integer $d > 1$, let $\mathcal{R}_{*, d} = \mathbb{Z}[X]/(X^d+1)$. Given an integer $q > 0$, let $\mathcal{R}_{q,d} = \mathbb{Z}_q[X]/(X^d+1) \simeq \mathcal{R}_{*, d} / q \mathcal{R}_{*, d}$. Polynomials are written in roman (e.g. $\mathrm{q}, \mathrm{k})$ and vectors are written in bold (e.g. $\mathbf{q}, \mathbf{k}$). Given a vector $\mathbf{v} \in F^t$, $v[i]$ denotes the $i$-th coordinate. Given polynomials $\mathrm{p}, \mathrm{p}' \in \mathcal{R}_{*, d}$, $\mathrm{p} \cdot \mathrm{p}' \in \mathcal{R}_{*,d}$ denotes the ring multiplication in $\mathcal{R}$. Given a polynomial $\mathrm{p} \in \mathcal{R}_{*,d}$, $\mathrm{p}[i]$ denotes the coefficient of $X^i$. Given a polynomial in $\mathrm{p} \in \mathcal{R}_{*, d'}$, we denote $\tilde{\mathrm{p}} \in \mathcal{R}_{*, d}$ as the natural embedding $\tilde{\mathrm{p}}(X) = \mathrm{p}(X^{d/d'})$. As we use $d$ as the fixed RLWE dimension, we omit $d$ in the notation $\tilde{\mathrm{p}}$.

The most significant difference between \textit{semantic search} and a \textit{vector database} is that the database must be dynamic, supporting insertion and deletion.  
An important observation is that HE operations should ideally not be used for insertion and deletion, as they accumulate errors and eventually corrupt the message.\footnote{One may consider using bootstrapping~\cite{Gen09} to clean the errors, but it is almost infeasible due to its high computational cost.}  
Many existing HE-based inner-product algorithms are unsuitable for scenarios requiring dynamic updates. Current solutions for \textit{encrypted semantic search} with a public database, such as Wally~\cite{wally} and HERS~\cite{hers}, typically precompute key vectors in plaintext domain for fast search. However, this plaintext precomputation restricts dynamic updates in the ciphertext domain. In HERS, for instance, each key data point is distributed across different ciphertexts, necessitating complex homomorphic encryption (HE) operations for inserting or deleting keys along with their approximate values. This process can degrade data integrity over time due to accumulated errors resulting from frequent HE computations.

One way to avoid HE computations during insertion and deletion is to assign one ciphertext per key, allowing insertion and deletion by simply appending or removing ciphertexts. 
We designed a dedicated HE-IP scheme for this scenario, achieving both \textbf{exact updates} and \textbf{fast search}.\footnote{CHAM~\cite{cham} also supports exact updates but is far less efficient than ours.}  

The search process begins by computing the inner product between the query and the stored key vectors.  
Let us break down each step to derive the complete algorithm.  
For simplicity, we first solve the case where \( n = d \) and \( r \) is a power of two.  
For \( n \geq d \), we can extend the base case to compute multiple similarity scores. We describe the behavior of the underlying plaintexts.

\paragraph{Inner Product.}  
Let the query vector be \( \mathbf{q} = [\xi_i]_{0 \leq i < r} \in \mathbb{R}^r \) and the key vector be \( \mathbf{k} = [\kappa_i]_{0 \leq i < r} \in \mathbb{R}^r \). The corresponding plaintext polynomials are encoded as 
\[\mathrm{q}(X) = \sum_{i=0}^{r-1} q_i \cdot X^{-si} = \sum_{i=0}^{r-1} \lfloor \Delta \cdot \xi_{i} \rceil \cdot X^{-si} \in \mathcal{R}_{*,d}\]
and
\[\quad \mathrm{k} (X) = \sum_{i=0}^{r-1} k_i \cdot X^{si} = \sum_{i=0}^{r-1} \lfloor \Delta \cdot \kappa_{i} \rceil \cdot X^{si} \in \mathcal{R}_{*, d},\] 
where $\Delta > 0$ is a scaling factor and \( d = rs \). Here the inner product $\langle \mathbf{q}, \mathbf{k} \rangle$ can be (approximately) derived as

\[ \frac{1}{\Delta^2} \cdot (\mathrm{q} \cdot \mathrm{k})[0] \simeq \langle \mathbf{q}, \mathbf{k} \rangle.
\]
We denote the (scaled) score $\sigma$ as $\sigma = (\mathrm{q} \cdot \mathrm{k})[0]$. To pack multiple scores in a ciphertext for reducing communication, we extract the constant term from the ciphertext. We slightly modify the conventional homomorphic trace and write 

\begin{equation}
\label{eq:single}
\sum_{i=0}^{r-1}\varphi_{i}(\mathrm{q} \cdot \mathrm{k}) = r \cdot \sigma
\end{equation}

where \( \varphi_{i} = \mathrm{p}(X) \mapsto \mathrm{p}(X^{2i+1})\) is an automorphism over $\mathcal{R}_{*,d}$ for each $0 \leq i < r$.

\paragraph{Batching.} We pack \( d \) scores $\sigma_0, \sigma_1, \ldots, \sigma_{d-1}$ into a single ciphertext. By Equation~\ref{eq:single},

\begin{equation}
\label{eq:batch}
r \cdot \sum_{j=0}^{d-1}\sigma_{j}X^{j} = \sum_{j=0}^{d-1}\sum_{i=0}^{r-1}\varphi_{i}(\mathrm{q} \cdot \mathrm{k}_{j})X^{j} = \sum_{i=0}^{r-1} \left[ \varphi_{i}(\mathrm{q}) \cdot \left( \sum_{j=0}^{d-1} \varphi_{i}(\mathrm{k}_{j})X^{j} \right) \right]
\end{equation}

where \( \sigma_{j} = (\mathrm{q} \cdot \mathrm{k}_{j})[0] \) for each $0 \leq j < d$. Here we observe that the last term can be interpreted as an inner product between $(\varphi_i)_i$ and $\left( \sum_{j=0}^{d-1} \varphi_i(\mathrm{k}_j) X^j \right)_i$, separating query and key operations. The number of automorphisms for the query is independent of \( n \) (when $n \geq d$), and we can precompute (i.e., cache) the keys.

\paragraph{Caching.} The key observation is that from Equation~\ref{eq:batch},
\begin{equation*}
    \sum_{j=0}^{d-1} \varphi_{i}(\mathrm{k}_{j})X^{j} = \varphi_{i} \left( \sum_{j=0}^{d-1} \mathrm{k}_{j}X^{j \cdot \mathtt{inv}(i)} \right)
\end{equation*}

where \( \mathtt{inv}(i) = (2i+1)^{-1} \ \mathtt{mod} \ 2d \) so that \( \varphi_{i}(X^{\mathtt{inv}(i)}) = X \).
 
This formula allows us to compute the automorphism \( \varphi_{i} \) only once.  
Therefore, we can significantly reduce the number of (homomorphic) automorphisms from \( d \log(r) \) to \( r - 1 \).  

\paragraph{Butterfly Decomposition.} For \( \mathbf{\tilde{k}} = \left(\tilde{k}_{j}\right)_{0 \leq j < d } \in \mathcal{R}_{q,d}^d \) and \( \mathbf{k} = \left(\sum_{j=0}^{d-1} \tilde{k}_{j}X^{j \cdot \mathtt{inv}(i)}\right)_{0 \leq i < r } \in \mathcal{R}_{q,d}^r \), let

\[
\mathbf{M} = P \cdot
\begin{bmatrix}
X^{0} & X^{1} & X^{2} & \cdots & X^{(d-1)} \\
X^{0} & X^{3} & X^{6} & \cdots & X^{3(d-1)} \\
X^{0} & X^{5} & X^{10} & \cdots & X^{5(d-1)} \\
\vdots & \vdots & \vdots & \ddots & \vdots \\
X^{0} & X^{2r-1} & X^{2(2r-1)} & \cdots & X^{(2r-1)(d-1)\}} 
\end{bmatrix} \in \mathcal{R}_{q,d}^{r \times d}
\]
where \( P \in \mathcal{R}_{q,d}^{r \times r} \) is a permutation matrix that corresponds to the permutation \( i \mapsto \frac{(2i+1)^{-1}-1}{2} \) mod \( r : \{0, 1, \ldots, r-1\} \rightarrow \{0, 1, \ldots, r-1\} \). Then \( \mathbf{k} = \mathbf{M}\mathbf{\tilde{k}} \) holds. 

Multiplying \( \mathbf{M} \) to \( \mathbf{\tilde{k}} \) requires \( r(r-1) \) polynomial additions, which is not negligible. Therefore, we use a DFT-style butterfly decomposition to reduce the computational cost.

Define \( \mathbf{k}' \in \mathcal{R}_{q,d}^{r} \) as follows:

\begin{equation*}
   \mathbf{k}'[i] = \sum_{j=0}^{s-1}\tilde{k}_{j+si}X^{j}
\end{equation*}

for \( 0 \leq i < r \). Then for \( \mathbf{k}'' = \mathbf{B}\mathbf{k} \in \mathcal{R}_{q,d}^{r} \), 
\[
\varphi_{i,r}(\mathbf{k}''[i]) = \varphi_{i}(\mathbf{k}[i])
\]
holds for \( 0 \leq i < r \), where

\begin{equation}
\label{eq:defB}
\mathbf{B} = P \cdot
\begin{bmatrix}
X^{0} & X^{s} & X^{2s} & \cdots & X^{(d-s)} \\
X^{0} & X^{3s} & X^{6s} & \cdots & X^{3(d-s)} \\
X^{0} & X^{5s} & X^{10s} & \cdots & X^{5(d-s)} \\
\vdots & \vdots & \vdots & \ddots & \vdots \\
X^{0} & X^{s(2r-1)} & X^{2s(2r-1)} & \cdots & X^{(2r-1)(d-s)\}} 
\end{bmatrix} \in \mathcal{R}_{q,d}^{r \times r}
\end{equation}
and $\varphi_{i,r} : \mathcal{R}_{q,d} \rightarrow \mathcal{R}_{q,d}$ is a permutation on the coefficients that satisfies
\[ \varphi_{i,r}(\mathrm{p})[(2i+1) \cdot s \cdot u + j] = \mathrm{p}[s \cdot u + j] \]
for $0 \leq j < s-1$ and $0 \leq u < r$. By leveraging the butterfly matrix decomposition, we reduce the number of polynomial additions to \( r\log(r) \). The detailed algorithm is written in~\Cref{alg:cache}.

\paragraph{Removing the Leading Term \( r \).} To remove the leading term \( r \) from the result \( r \cdot \sum_{j=0}^{d-1}\sigma_{j}X^{j} \), we multiply \( r^{-1} \) (mod \( q \)) before automorphisms.

\begin{equation*}
   r \cdot \sum_{i=0}^{r-1} \left( \varphi_{i}(r^{-1} \cdot \mathrm{q}) \cdot \left[ r \cdot \varphi_{i} \left( \sum_{j=0}^{d-1} r^{-1} \cdot \mathrm{k}_{j} X^{j \cdot \mathtt{inv}(i)} \right) \right] \right) = r \cdot \sum_{j=0}^{d-1}\sigma_{j}X^{j}
\end{equation*}

Therefore, \( \sum_{i=0}^{r-1} \left( \varphi_{i}(r^{-1} \cdot \mathrm{q}) \cdot \left[ r \cdot \varphi_{i} \left( \sum_{j=0}^{d-1} r^{-1} \cdot \mathrm{k}_{j} X^{j \cdot \mathtt{inv}(i)} \right) \right] \right) = \sum_{j=0}^{d-1}\sigma_{j}X^{j} \).

\paragraph{Optimizations.}
To enhance the performance of homomorphically encrypted vector databases, we incorporate several advanced techniques to optimize computation, storage, and accuracy.
One key optimization is caching via key-query decoupling, which allows keys to be precomputed and cached independently of queries.
This significantly reduces query response time by accelerating inner product computation.
We also apply hoisting~\cite{helib-lintrans, hoist} to efficiently decompose queries, minimizing computational overhead.
This technique, when combined with MLWE (Module Learning With Errors)~\cite{hermes} and seed-based ciphertext generation, enables compact storage and efficient updates.
Storage and update efficiency is further improved through batch processing and MLWE-based seeding strategies~\cite{kyber}, which reduce ciphertext size and update costs.
Finally, we improve numerical precision by removing leading constant terms~\cite{cdks} in homomorphic computations, resulting in more accurate query results.
See Appendix~\ref{sec:appendix_database} for detailed descriptions, and Section~\ref{sec:experiments_database} for results on latency, storage, and accuracy.

\subsection{Database Operations}
\label{subsec:encrypted_vector_db_op}

See Table~\ref{tab:algorithms} for our database’s API. With these APIs (functions), we achieve efficient search and support dynamic updates with \( O(1) \) complexity.

\begin{table*}[h]
  \centering
  \begin{minipage}[t]{0.48\textwidth}
    \centering
    \input{algorithms/init}
    \vspace{-2em}
    \input{algorithms/search}
    \vspace{-2em}
    \input{algorithms/retrieve}
  \end{minipage}%
  \hfill
  \begin{minipage}[t]{0.48\textwidth}
    \centering
    \input{algorithms/insert}
    \vspace{-2em}
    \input{algorithms/delete}
  \end{minipage}
  \caption{
  Set of algorithms for homomorphically encrypted vector database operations.
  }
  \label{tab:algorithms}
\end{table*}

These operations include initialization, encrypted search and retrieval, as well as insertion and deletion. We denote the client as \textbf{Alice (A)} and the server as \textbf{Bob (B)}. Public parameters \texttt{pp} are shared between them. The function \texttt{GenSK} generates secret keys used in Homomorphic Encryption (HE), Advanced Encryption Standard (AES), and Private Information Retrieval (PIR), while \texttt{GenSwk} produces the corresponding public keys of HE and PIR including switching keys for homomorphic operations.

The finite-length sequences, such as textual queries \( q \) and records \( v \) , are embedded using an encoder \(E\). The vector database \( \mathcal{D} \) maintains the following attributes: \texttt{num} (number of entries), \texttt{key} (stored key vectors), \texttt{value} (encrypted records), and \texttt{cache} (cached key vectors for efficient search). Encryption and decryption are performed using \texttt{EncryptHE}, \texttt{DecryptHE}, \texttt{EncryptAES}, and \texttt{DecryptAES}.

The \texttt{Score} function computes similarity scores over encrypted vectors, and \texttt{TopK} selects the top-\(k\) most relevant entries using a heap-based algorithm with \( O(n \log k) \) complexity. Retrieved values are fetched securely using PIR protocols. See Appendix~\ref{sec:appendix_database} for more details.

To support dynamic updates, we include auxiliary operations such as \texttt{len} (entry count), \texttt{Append} (inserting new entries), \texttt{Switch} (deleting entries by overwriting them with the last entry), and \texttt{ReCache} (refreshing the cached key vectors). These operations are executed in a batched manner and achieve constant-time complexity.

\paragraph{Security Guarantees.} 
Key vectors are encrypted using CKKS~\cite{ckks}.
Values in the vector database are encrypted using non-deterministic AES-256 encryption.  
The combination of HE and AES provides robust security of our vector database. That is, our database provides 128-bit IND-CPA security~\cite{aesind, security-guidelines} and is quantum-resistant~\cite{bonnetain2019quantum, latticepqc}. 

%% file: algorithms/init.tex
\begin{algorithm}[H]
\caption{\texttt{Init}}
\begin{algorithmic}[1]
    \REQUIRE public parameters \( \mathtt{pp} \)
    \STATE \textbf{A:} \( \mathtt{sk} \gets \mathtt{GenSK}(\mathtt{pp}) \)
    \STATE \textbf{A:} \( \mathtt{pk} \gets \mathtt{GenPK}(\mathtt{pp}) \)
    \STATE \textbf{A:} Send \( \mathtt{pk} \) to \textbf{Bob}
\end{algorithmic}
\end{algorithm}

%% file: algorithms/search.tex
\begin{algorithm}[H]
\caption{\texttt{Search}}
\begin{algorithmic}[1]
    \REQUIRE query \( q \), database \( \mathcal{D} \)
    \STATE \textbf{A:} \( \mathbf{q} \gets E(q) \)
    \STATE \textbf{A:} \( \mathsf{q} \gets \mathtt{EncryptHE}(\mathbf{q}) \)
    \STATE \textbf{A:} Send \( \mathsf{q} \) to \textbf{Bob}
    \STATE \textbf{B:} \( \mathsf{s} \gets \mathtt{Score}(\mathsf{q}, \mathcal{D}_{\mathtt{cache}}) \)
    \STATE \textbf{B:} Send \( \mathsf{s} \) to \textbf{Alice}
    \STATE \textbf{A:} \( \mathbf{s} \gets \mathtt{DecryptHE}(\mathsf{s}) \)
    \STATE \textbf{A:} \( \mathcal{I} \gets \mathtt{TopK}(\mathbf{s}) \)
\end{algorithmic}
\end{algorithm}

%% file: algorithms/retrieve.tex
\begin{algorithm}[H]
\caption{\texttt{Return}}
\begin{algorithmic}[1]
    \REQUIRE record ids \( \mathcal{I} \)
    \STATE \textbf{A\&B:} \( \{v\} \gets  \mathtt{PIR}(\mathcal{D}_{\mathtt{value}}, \mathcal{I}) \)
    \STATE \textbf{A:} \( \{v\} \gets \{\mathtt{DecryptAES}(v)\} \)
\end{algorithmic}
\end{algorithm}

%% file: algorithms/insert.tex
\begin{algorithm}[H]
\caption{\texttt{Insert}}
\begin{algorithmic}[1]
    \REQUIRE set of records \( \{ v \} \), database \( \mathcal{D} \)
    \STATE \textbf{A:} \( \mathbf{k} \gets E(v) \)
    \STATE \textbf{A:} \( \mathsf{k} \gets \mathtt{EncryptHE}(\mathbf{k}) \)
    \STATE \textbf{A:} \( v \gets \mathtt{EncryptAES}(v) \)
    \STATE \textbf{A:} Send \( \{ (\mathsf{k}, v) \} \) to \textbf{Bob}  
    \STATE \textbf{B:} \( \mathcal{D}_{\mathtt{num}} \gets \mathcal{D}_{\mathtt{num}} + \mathtt{len}( \{ (\mathsf{k}, v) \} ) \)
    \STATE \textbf{B:} \( \mathcal{D}_{\mathtt{key}} \gets \mathtt{Append}(\mathcal{D}_{\mathtt{key}}, \{ \mathsf{k} \} ) \)
    \STATE \textbf{B:} \( \mathcal{D}_{\mathtt{value}} \gets \mathtt{Append}(\mathcal{D}_{\mathtt{value}}, \{ v \} ) \)
    \STATE \textbf{B:} \( \mathcal{D}_{\mathtt{cache}} \gets \mathtt{ReCache}(\mathcal{D}_{\mathtt{cache}},  \mathcal{D}_{\mathtt{key}}, \{ \mathsf{k} \} ) \)
\end{algorithmic}
\end{algorithm}

%% file: algorithms/delete.tex
\begin{algorithm}[H]
\caption{\texttt{Delete}}
\begin{algorithmic}[1]
    \REQUIRE record ids \( \mathcal{A} \), database \( \mathcal{D} \)
    \STATE \textbf{A:} Send \( \mathcal{A} \) to \textbf{Bob}
    \STATE \textbf{B:} \( \mathcal{D}_{\mathtt{num}} \gets \mathcal{D}_{\mathtt{num}} - \mathtt{len}(\mathcal{A}) \)
    \STATE \textbf{B:} \(  \mathcal{D}_{\mathtt{key}} \gets \mathtt{Switch}(\mathcal{D}_{\mathtt{key}}, \mathcal{A} ) \)
    \STATE \textbf{B:} \( \mathcal{D}_{\mathtt{value}} \gets \mathtt{Switch}(\mathcal{D}_{\mathtt{value}}, \mathcal{A} ) \)
    \STATE \textbf{B:} \( \mathcal{D}_{\mathtt{cache}} \gets \mathtt{ReCache}(\mathcal{D}_{\mathtt{cache}}, \mathcal{D}_{\mathtt{key}}, \mathcal{A} ) \)
\end{algorithmic}
\end{algorithm}

%% file: sections/5_experiment.tex
\section{Experiments}
\label{sec:experiments}

In this section, we empirically validate the effectiveness of our privacy-preserving framework.  
The experiments are organized into three parts.  
We first present the overall performance of the full framework.  
We then conduct ablations on Socratic Chain-of-Thought Reasoning, isolating the contributions of sub-query generation and chain-of-thought generation.  
Finally, we examine the accuracy of our encrypted database and evaluate its efficiency and scalability in terms of latency and storage cost.

\subsection{Main Results}
Experiments are conducted on two question-answering benchmarks: LoCoMo~\cite{locomo}, designed to simulate personal assistant scenarios, and MediQ~\cite{mediq}, aimed at simulating medical consultation scenarios. Both tasks require retrieving relevant user-specific data and performing complex reasoning to generate an accurate final answer. We use DRAGON~\cite{dragon} to obtain embedding vectors, facilitating the retrieval of proper records related to each query. See Appendix~\ref{sec:appendix_experimental_setup} for more details on the experimental setup. Consequently, experiments evaluate whether the model's final responses, derived from the user's original query and stored personal data, align closely with the desired answers.

\input{tables/4_experiment_benchmark}
 
\paragraph{Our framework improves local-only baselines by up to +27.6 percentage points.}
As shown in Table~\ref{tab:4_experiment_benchmark}, our framework consistently outperforms the local-only baselines on both the LoCoMo and MediQ datasets.
By delegating complex reasoning to powerful remote models, we observe substantial gains in performance.
Specifically, we see improvements of up to 23.1 percentage points on LoCoMo and 27.6 on MediQ when comparing each local model with its corresponding privacy-preserving variants.
On average, our approach improves F1 by +19.8 percentage points on LoCoMo and exact match by +19.0 percentage points on MediQ over the local-only counterparts.
These gains are especially notable in challenging scenarios requiring domain expertise, such as medical consultations.
Despite operating under strict privacy constraints, our framework approaches—and in some cases surpasses—the performance of oracle baselines that operate without privacy constraints.
This demonstrates the effectiveness of our approach in balancing strong privacy with high utility.

\subsection{Ablations on Socratic Chain-of-Thought Reasoning}

To better understand the source of performance gains from Socratic Chain-of-Thought Reasoning, we conduct two ablation studies on the LoCoMo~\cite{locomo} and MediQ~\cite{mediq} datasets.

\paragraph{Reasoning augmentation leads to substantial performance gains.}
Table~\ref{tab:4_experiment_ablation_1} compares remote-only and local-only baselines, with and without Socratic Chain-of-Thought Reasoning.  
On LoCoMo, all methods benefit from reasoning augmentation: explicitly prompting the model to reason through intermediate steps leads to clear performance gains.  
For example, the local-only baseline improves from 64.6 to 82.0, a gain of +17.4 percentage points, while the remote-only baseline improves from 80.6 to 92.6, a gain of +12.0 percentage points.  
These results suggest that reasoning augmentation through Socratic Chain-of-Thought Reasoning is key to performance gains on LoCoMo.

\input{tables/4_experiment_ablation_1}

\paragraph{Delegating both sub-queries and chain-of-thought generation to more powerful models is key.}
Table~\ref{tab:4_experiment_ablation_2} highlights two key observations by isolating the contributions of sub-query generation and chain-of-thought generation.  

\emph{First, delegating sub-query generation significantly improves retrieval quality.}
On LoCoMo, using a smaller model (Llama-3.2-1B) for sub-query generation limits retrieval performance (Recall@5 = 21.8).  
When this task is handled by a more capable model (GPT-4o), performance nearly doubles to 44.1.  
This indicates that sub-query generation often requires deeper understanding and reasoning, which smaller models struggle to achieve.  
Furthermore, using ground-truth retrieval results boosts performance even more, implying that better sub-query generation—closer to the ideal target—can further enhance final answer quality.  
On MediQ, the amount of private data per user is so limited that most of the relevant records are retrieved even without high-quality sub-queries, reducing the impact of sub-query generation on overall performance.  

\emph{Second, delegating chain-of-thought generation improves final response quality.}
On LoCoMo, without any chain-of-thought (N/A), the F1 score is 77.8. 
Incorporating chain-of-thought reasoning from the smaller model raises it to 85.4, and using GPT-4o improves it further to 89.3.  
These results demonstrate that guiding generation with reasoning augmentation produced by stronger models plays a critical role in achieving high answer quality.  
Meanwhile, on MediQ, augmenting reasoning without rich domain knowledge from remote models yields only marginal improvements.  
In this case, the dominant factor is qualified reasoning criteria generated with rich domain knowledge, which powerful remote models provide far more effectively than smaller local models.
We provide a more detailed analysis of the MediQ results in Appendix~\ref{sec:appendix_additional_mediq_analysis}.

\input{tables/4_experiment_ablation_2}

These findings suggest that local-only baselines, even without disclosing queries, are sufficient as effective personal assistants for casual tasks like LoCoMo. In contrast, specialized domains such as MediQ necessitate leveraging the advanced expertise embedded within powerful remote models to deliver high-quality answers.
\emph{Therefore, collaborating with remote models becomes essential for users seeking more accurate responses in expert domains.}

\subsection{Homomorphically Encrypted Vector Database}
\label{sec:experiments_database}

\paragraph{Encrypted search retains > 99\% accuracy.}
We evaluate the search accuracy of our encrypted database using benchmarks from LoCoMo~\cite{locomo}, Deep1B~\cite{deep1b}, and LAION~\cite{laion}, covering a range of vector dimensions and domains.
Results in Table~\ref{tab:4_experiment_search_accuracy} show that the system maintains high search fidelity across both plaintext-to-ciphertext and ciphertext-to-ciphertext inner product computations.
In particular, when the query is in plaintext—a setting aligned with our privacy-preserving framework involving non-private queries and private data—the encrypted database achieves accuracy comparable to its unencrypted counterpart, with both mean and maximum inner product errors remaining minimal.
Metrics such as 1-Recall@1, 1-Recall@5, and MRR@10 confirm that the top-\(k\) results from the encrypted database closely mirror those of the plaintext system.
These results demonstrate that encrypted search can be performed with negligible impact on accuracy.

\input{tables/4_experiment_search_accuracy}

\paragraph{Encrypted search scales to 1M entries with < 1 second latency.}
Despite the typical computational overhead of homomorphic encryption, our system achieves practical latency for large-scale vector similarity search.
Figure~\ref{fig:deep1b} presents results on the Deep1B~\cite{deep1b} dataset, showing that by leveraging efficient SIMD-style operations and low-precision arithmetic in ciphertext space, the system achieves linear scalability across database sizes from 1{,}000 to 1 million entries.
\emph{Even at the million scale, end-to-end latency remains under one second, including encryption, computation, and communication—even under a slow network.}
This performance makes the encrypted database viable for real-time applications.
Network overhead has become the primary source of latency, reflecting that computation—particularly homomorphic encryption—no longer constitutes the principal bottleneck.
This improvement is evident in our evaluation on the 100K subset of LAION dataset~\cite{laion}, where encrypted search completes in 62\,ms on the fast network and 251\,ms on the slow network, compared to 76\,ms and 931\,ms with Compass~\cite{compass}---yielding \(1.2\times\) and \(3.7\times\) speed-ups, respectively.

\input{figures/deep1b}

\input{tables/4_experiment_search_latency}

In addition, our system achieves significant improvements over previous homomorphic encryption methods.
As shown in Table~\ref{tab:4_experiment_search_latency}, our approach consistently delivers faster runtimes than CHAM~\cite{cham}, an encrypted matrix–vector multiplication method designed to support frequent updates.
The performance gap widens with scale: while CHAM requires 84{,}543\,ms to process 1 million entries, our method completes the same operation in just 2{,}280\,ms—achieving a \( \mathbf{37\times} \) speed-up.
This efficiency primarily stems from our query caching strategy, which restructures the key-switching phase so that its computational complexity scales with the vector length rather than the full matrix size, effectively eliminating the dominant bottleneck in prior designs.

\paragraph{Encrypted storage incurs < \( \mathbf{5.8\times} \) overhead.}
Storing high-dimensional vectors in homomorphic ciphertexts introduces nontrivial storage overhead.
However, as detailed in Section~\ref{sec:database} and Appendix~\ref{sec:appendix_database}, our implementation adopts optimizations such as packing multiple vector components into a single ciphertext and omitting unused polynomial coefficients, effectively reducing space requirements.
Moreover, we apply module-LWE variants and seed-based ciphertext generation techniques, which scale ciphertext size \emph{linearly} with vector dimensionality rather than polynomial degree.
As a result, the encrypted database achieves practical storage costs, less than \( \mathbf{5.8\times} \) overhead even for millions of entries, enabling deployment in real-world systems without requiring excessive disk resources.

%% file: tables/4_experiment_benchmark.tex
\begin{table*}[h]
\centering
\begin{tabular}{@{}llcc@{}}
\toprule
\textbf{Baseline} & \textbf{Model} & \textbf{LoCoMo} & \textbf{MediQ} \\
\midrule
\multirow{3}{*}{\textbf{Remote-Only Baseline} (oracle)} 
& R1 \space GPT-4o & 80.6 & \textbf{81.8} \\
& R2 \space Gemini-1.5-Pro & 84.2 & 69.8 \\
& R3 \space Claude-3.5-Sonnet & \textbf{89.8} & 79.3 \\
\midrule
\multirow{3}{*}{\textbf{Local-Only Baseline}} 
& L1 \space Llama-3.2-1B & 64.6 & 32.1 \\
& L2 \space Llama-3.2-3B & 68.7 & 43.2 \\
& L3 \space Llama-3.1-8B & 68.8 & 47.5 \\
\midrule
\multirow{9}{*}{\textbf{Hybrid Framework w/ Socratic CoT} (ours)} 
& L1 + R1 & 87.7 & 59.7 \\
& L1 + R2 & 85.1 & 49.7 \\
& L1 + R3 & 84.3 & 58.0 \\
\cdashline{2-4}
\noalign{\vskip 2.5pt}
& L2 + R1 & 85.9 & 60.7 \\
& L2 + R2 & 79.8 & 52.9 \\
& L2 + R3 & 74.6 & 59.0 \\
\cdashline{2-4}
\noalign{\vskip 2.5pt}
& L3 + R1 & 87.9 & 59.5 \\
& L3 + R2 & 88.0 & 52.1 \\
& L3 + R3 & 86.1 & 59.6 \\
\bottomrule
\end{tabular}
\vspace{1.0em}
\caption{
Benchmark results on the \textbf{LoCoMo} and \textbf{MediQ} datasets.
LoCoMo is evaluated by F1 score, while MediQ is evaluated by exact match.
\emph{Takeaway: Our privacy-preserving framework significantly outperforms local-only baselines and approaches the performance of oracle baselines without privacy constraints.}
\minchan{I will add more results.}
}
\label{tab:4_experiment_benchmark}
\end{table*}

%% file: tables/4_experiment_ablation_1.tex
\begin{table}[h]
\centering
\begin{tabular}{lccc}
\toprule
\textbf{Method} & \textbf{Model} & \textbf{LoCoMo} & \textbf{MediQ} \\
\midrule
\textbf{Remote-Only Baseline} & R1 & 80.6 & \textbf{81.8} \\
\textbf{Remote-Only Baseline w/ Socratic CoT} & R1 + R1 & \textbf{92.6} & 67.3 \\
\midrule
\textbf{Local-Only Baseline} & L1 & 64.6 & 32.1 \\
\textbf{Local-Only Baseline w/ Socratic CoT} & L1 + L1 & 82.0 & 32.5 \\
\midrule
\textbf{Hybrid Framwork w/ Socratic CoT} (ours) & L1 + R1 & 87.7 & 59.7 \\
\bottomrule
\end{tabular}
\vspace{1.0em}
\caption{
The first ablation study for Socratic Chain-of-Thought Reasoning on the \textbf{LoCoMo} and \textbf{MediQ} datasets.
LocoMo is evaluated by F1 score, while MediQ is evaluated by exact match.
R1 is GPT-4o, and L1 is Llama-3.2-1B.
\emph{Takeaway: Reasoning augmentation through Socratic Chain-of-Thought Reasoning is the primary driver of performance gains.}
}
\label{tab:4_experiment_ablation_1}
\end{table}

%% file: tables/4_experiment_ablation_2.tex
\begin{table}[h]
\centering
\begin{minipage}[t]{0.48\textwidth}
\centering
\begin{tabular}{l|ccc}
\toprule
\diagbox[width=10em, height=2em]{\textbf{Sub-Query}}{\textbf{CoT}} & \textbf{R1} & \textbf{L1} & \textbf{N/A} \\
\midrule
\textbf{GT} & 89.3 & 85.4 & 77.8 \\
\textbf{R1} (GPT-4o) & 87.7 & 84.7 & 73.9 \\
\textbf{L1} (Llama-3.2-1B) & 84.9 & 82.0 & 64.6 \\
\bottomrule
\end{tabular}
\vspace{0.5em}
\caption*{\textbf{(a) LoCoMo}}
\end{minipage}
\hfill
\begin{minipage}[t]{0.48\textwidth}
\centering
\begin{tabular}{l|ccc}
\toprule
\diagbox[width=10em, height=2em]{\textbf{Sub-Query}}{\textbf{CoT}} & \textbf{R1} & \textbf{L1} & \textbf{N/A} \\
\midrule
\textbf{All} & 60.4 & 32.1 & 31.4 \\
\textbf{R1} (GPT-4o) & 59.7 & 31.8 & 33.2 \\
\textbf{L1} (Llama-3.2-1B) & 58.6 & 32.5 & 32.0 \\
\bottomrule
\end{tabular}
\vspace{0.5em}
\caption*{\textbf{(b) MediQ}}
\end{minipage}
\caption{
The second ablation study for Socratic Chain-of-Thought Reasoning on the \textbf{LoCoMo} and \textbf{MediQ} datasets.
LocoMo is evaluated by F1 score, while MediQ is evaluated by exact match.
Each row corresponds to a different sub-query generation method:
For LoCoMo, GT uses ground-truth private data without sub-query generation (Recall@5=\textbf{100.0}), R1 uses GPT-4o (Recall@5=\textbf{44.1}), and L1 uses Llama-3.2-1B (Recall@5=\textbf{21.8}).
For Mediq, All setup uses the full user history as input since no retrieval annotation is available, while R1 and L1 follow the same retrieval configuration as in LoCoMo.
Each column corresponds to a different chain-of-thought generation method, where N/A indicates that chain-of-thought reasoning is not used.
L1 is used for final response generation across all settings.
\emph{Takeaway: Delegating both sub-query and chain-of-thought generation to more powerful models is crucial for optimal performance.}
}
\label{tab:4_experiment_ablation_2}
\end{table}

%% file: tables/4_experiment_search_accuracy.tex
\begin{table}[h]
\centering
\begin{tabular}{@{}lcccccc@{}}
\toprule
\textbf{Dataset} & \textbf{Max Error} & \textbf{Mean Error} & \textbf{Std Error} & \textbf{MRR@10} & \textbf{1-Recall@1} & \textbf{1-Recall@5} \\
\midrule
\multicolumn{7}{l}{\textbf{Plaintext Query}} \\ 
\midrule
\textbf{LoCoMo} & 3.11e-3 & 2.97e-3 & 3.31e-9 & 99.99 & 99.97 & 100 \\
\textbf{Deep1B} & 5.29e-5 & 6.42e-6 & 7.00e-11 & 99.97 & 99.96 & 99.99 \\
\textbf{LAION} & 1.06e-4 & 9.83e-6 & 1.36e-10 & 99.86 & 99.79 & 99.95 \\
\midrule
\multicolumn{7}{l}{\textbf{Ciphertext Query}} \\
\midrule
\textbf{LoCoMo} & 5.31e-2 & 2.32e-2 & 2.31e-1 & 93.31 & 89.20 & 98.89 \\
\textbf{Deep1B} & 1.39e-3 & 1.71e-4 & 4.61e-8 & 99.59 & 99.21 & 99.97 \\
\textbf{LAION} & 2.70e-3 & 3.44e-4 & 1.87e-7 & 99.85 & 99.78 & 99.95 \\
\bottomrule
\end{tabular}
\vspace{1.0em}
\caption{
Search accuracy across \textbf{LoCoMo}~\cite{locomo}, \textbf{Deep1B}~\cite{deep1b}, and \textbf{LAION}~\cite{laion} datasets, evaluated under two settings: when the query is in plaintext (top) and when the query is encrypted (bottom), with encrypted keys in both cases.
\emph{Takeaway: Our encrypted database preserves high retrieval accuracy, achieving near-parity with the fully plaintext setting (both query and key).}
}
\label{tab:4_experiment_search_accuracy}
\end{table}

%% file: figures/deep1b.tex
\begin{figure}[h]
\centering
\includegraphics[width=0.8\textwidth]
{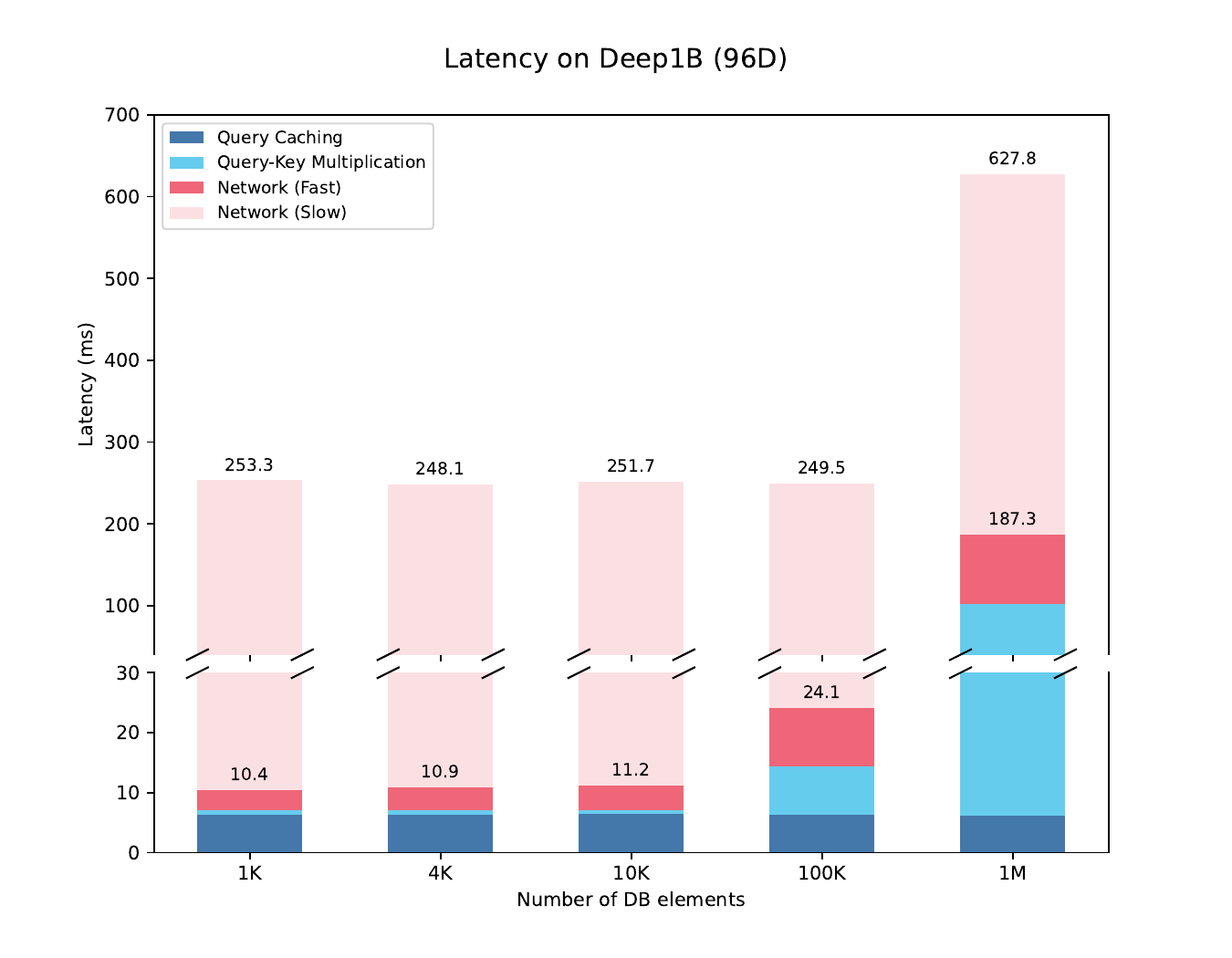}
\caption{
Multi-thread search latency (using 64 threads) breakdown on the Deep1B~\cite{deep1b} dataset as the number of database entries increases.
Red and pink bars represent network communication time on fast and slow networks, respectively, while the numbers above each bar indicate the corresponding latency.
Blue bars represent query caching time; light-blue bars show query-key multiplication time.
\emph{Takeaway: Our encrypted search scales to 1M entries with < 1 second latency, as homomorphic operations incur relatively low overhead compared to network communication.}
}

\label{fig:deep1b}
\end{figure}

%% file: tables/4_experiment_search_latency.tex
\begin{table}[h]
\centering
\begin{tabular}{lrrrrr}
        \toprule
        \textbf{Deep1B} & \textbf{1K} & \textbf{4K} & \textbf{10K} & \textbf{100K} & \textbf{1M} \\
        \midrule
        CHAM~\cite{cham} & 378 ms & 389 ms & 1,171 ms & 9,406 ms & 84,543 ms \\
        Ours & 150 ms & 151 ms & 156 ms & 236 ms & 951 ms \\
        \bottomrule
    \end{tabular}
\vspace{1.0em}
\caption{
Single-thread runtime of homomorphically encrypted matrix–vector multiplication as the number of vectors in the database increases.
The CHAM~\cite{cham} baseline is based on our re-implementation of the original method, incorporating additional optimizations such as ring packing and packing multiple vectors into a single ciphertext.
\emph{Takeaway: Our method achieves up to \( \mathbf{88\times} \) speed-up over CHAM, enabling real-time encrypted search at million-scale.}
}
\label{tab:4_experiment_search_latency}
\end{table}

%% file: sections/6_related_work.tex
\section{Related Work}

In this paper, we address the challenge of privacy-preserving LLM interaction, focusing on protecting user records in the context, at inference time. Unlike private training approaches which safeguard the training corpus through techniques like DP-SGD and DP-ICL~\citep{Abadi2016,tang2023privacy}, we focus on protecting user data provided as inputs to the model during inference, ensuring that sensitive context information remains confidential and is not leaked or memorized by the remote LLMs. Our work intersects with the following topics:

\noindent\textbf{Private Inference via Encryption.}
Early approaches combined homomorphic encryption (HE) with neural networks, exemplified by CryptoNets~\citep{GiladBachrach2016}, though with $10^3\times$ computational overhead. Subsequent systems like Gazelle~\citep{Juvekar2018} and XONN~\citep{Riazi2019} reduced latency by hybridizing HE with garbled circuits and binary networks. Recent work extends these techniques to Transformers and LLMs: MPCFormer~\citep{Mao2022}, PermLLM~\citep{Zhu2024}, and PUMA~\citep{Li2023} achieve privacy for BERT and LLaMA architectures but still require seconds per token. Industry implementations like Apple's HE+PIR photo search~\citep{Apple2024} show promise, but cloud LLM providers have been reluctant to adopt these approaches due to significant computational overhead and complex key management.

\noindent\textbf{Input Minimization and Sanitization Methods.}
Complementary approaches focus on sanitizing prompts before transmission. PREEMPT~\citep{chowdhury2025pr} detects and replaces sensitive spans with placeholders or differentially private values. PAPILLON~\citep{papillon} divides processing between local lightweight models and external LLMs, sending only abstracted prompts to the cloud. Additional work~\cite{dou2023reducing,staab2024large} focuses on abstracting personal information. While effective for specific domains, these approaches typically require task-specific engineering or sacrifice accuracy when critical context is removed~\cite{xin2025false}. Our framework preserves task performance without cryptographic overhead by keeping raw data in the trusted zone while delegating only non-sensitive reasoning steps.

\noindent\textbf{Chain-of-Thought Reasoning and Task Decomposition.}
Chain-of-thought (CoT) prompting has emerged as a powerful technique for improving LLM reasoning through step-by-step solutions. Zero-shot CoT techniques~\citep{Wei2022,Kojima2022} and task decomposition prompts~\citep{Zhou2022,Press2022} guide models to break complex problems into manageable sub-problems, often enhanced with supervised reasoning traces~\citep{Zelikman2022}. Parallel work on model cascades aims to maximize efficiency by routing queries between different-sized models, as in FrugalGPT~\citep{Chen2023Frugal} and Hybrid LLM~\citep{Ding2024}, typically using confidence estimators to determine when smaller models are insufficient~\citep{Liu2024OptLLM,Gupta2024}. Multi-model frameworks like Socratic Models~\citep{Zeng2022} and HuggingGPT~\citep{Shen2023} divide tasks between a powerful LLM planner and specialized executors, but assume the central model has full access to private data. In contrast, our approach performs test-time CoT decomposition without additional training while preserving privacy by ensuring the large LLM only sees abstracted queries rather than raw private data.

\noindent\textbf{RAG and Agentic Workflows.}
Recent systems increasingly embed LLMs within persistent user-centric datastores to deliver personalized assistance. These range from research prototypes like Generative Agents~\citep{Park2023} that maintain interaction histories as long-term memory, to commercial deployments such as ChatGPT's "Memory" and Operator~\citep{OpenAI2024Memory,OpenAI2024Operator} that preserve multi-day conversation logs, and open frameworks like LangChain and LlamaIndex~\citep{Mavroudis2024,LlamaIndex2024} that provide memory backends as first-class primitives. Life-logging assistants like Rewind and Lindy~\citep{RewindAI2024,LindyAI2024} index users' entire digital traces, leveraging the success of retrieval-augmented generation (RAG)~\citep{Lewis2020} for grounding LLMs in external knowledge. However, these systems typically assume trustworthy datastores, ignoring privacy risks highlighted by recent extraction and inference attacks~\citep{Bianchi2023}. Our framework is the first to combine an agentic RAG architecture with encrypted, local retrieval, addressing this critical privacy gap while maintaining the benefits of contextual personalization.

%% file: sections/7_conclusion.tex
\section{Conclusion and Discussion}
\label{sec:conclusion}

We introduced a four-stage, privacy-preserving framework that uniquely partitions tasks between untrusted powerful LLMs and trusted lightweight local models. Our key innovations—Socratic Chain-of-Thought Reasoning and Homomorphically Encrypted Vector Database—enable secure collaboration without exposing private data. Our approach not only preserves privacy but actually improves performance, with our local lightweight model outperforming even GPT-4o on long-context QA tasks. This counter-intuitive result demonstrates the power of additional test-time computation when properly structured through our chain-of-thought decomposition. Meanwhile, our encrypted vector database achieves sub-second latency on million-scale collections with negligible accuracy loss compared to plaintext search.

Future work should address extending our approach to tasks resistant to clean decomposition, developing dynamic sensitivity classification for mixed public-private content, and scaling encrypted retrieval to billion-scale collections. These advances will further expand applications that can benefit from powerful models without surrendering personal data.

%% file: sections/8_acknowledgments.tex
\begin{ack}
This study was supported by the NAVER Digital Bio Innovation Research Fund, funded by NAVER Corporation (Grant No.~3720230180).
Jaehyung Kim acknowledges support from the Stanford Graduate Fellowship, NSF, DARPA, and the Simons Foundation. Opinions, findings, and conclusions or recommendations expressed in this material are those of the authors and do not necessarily reflect the views of DARPA.
\end{ack}

%% file: sections/A_database.tex
\section{Homomorphic Encryption based Inner Product}
\label{sec:appendix_database}

\subsection{Secure Inner Product, Algorithms and Optimizations}

We specify the detailed algorithms as follows. \Cref{alg:decomp,alg:cache} describe the precomputations for the query and key, respectively, as mentioned right after~\Cref{eq:batch}. \Cref{alg:score} describes the score computation algorithm starting from the precomputed query and cache ciphertexts.

\paragraph{Optimizations Summary.}
We summarize the optimizations mentioned in the previous subsection and discuss some additional optimizations.

\begin{itemize}
    \item \textbf{Batching and Caching}: We write the homomorphic inner product equation as in~\Cref{eq:batch}. This separates the precomputations for query and key, which are denoted as \texttt{Decompose} and \texttt{Cache}, respectively. This reduces the number of automorphisms from \( d \log(r) \) to \( r-1 \).
    \item \textbf{Butterfly Decomposition}: The key side precomputation is significant as it involves $O(r^2)$ polynomial additions. We leverage the butterfly decomposition to reduce the complexity from \( r(r-1) \) to \( r \log(r) \).
    \item \textbf{Seeding and MLWE}: In order to improve the storage size, we use Module LWE (MLWE)~\cite{mlwe} and Extendable Output-format Function (XOF) with a public seed. This reduces ciphertext size from \( 2d \) (i.e. two $\mathcal{R}_{q,d}$ elements) to \( r \) (i.e. one $\mathcal{R}_{q,r}$ element and a 128-bit public seed).
    \item \textbf{Remove the leading term} \( r \): We use the optimization technique introduced in~\cite{cdks} that evaluates the trace without the leading term $r$, thereby improving the precision. This technique is applied for Line 2 of~\Cref{alg:decomp} and Line 3 of~\Cref{alg:cache}.
    \item \textbf{Hoisting}~\cite{helib-lintrans}: We adapt the hoisting technique that lazily computes the homomorphic operations to improve efficiency. Our adaptaion is similar to the double hoisting algorithm in~\cite{hoist}. Hoisting appears in the following instances.
        \begin{itemize}
            \item Line~3 of~\Cref{alg:decomp}:
            For each index $0 \leq i < s$, $\mathtt{ModUp}(a_i)$ is computed only once. 
            \item Line~5,6 of~\Cref{alg:decomp}, Line~13,14 of~\Cref{alg:cache}: We $\mathtt{ModDown}$ after summation, reducing the number of $\mathtt{modDown}$ to $r$ per each $j$.
        \end{itemize}
    \item \textbf{Reducing NTT dimension}: In Line~3,5,6 of~\Cref{alg:decomp}, we utilize dimension $r$ NTT instead of dimension $d$ NTT, reducing the complexity by a factor of $\log(d)/\log(r)$. This is possible because each $\hat{a}_i$ is sparsely embedded into the larger ring $\mathcal{R}_{q,d}$.
\end{itemize}

\begin{algorithm}[H]
\caption{\texttt{Decompose}}
\label{alg:decomp}
\begin{algorithmic}[1]
    \REQUIRE Query (seeded) MLWE ciphertext \( ( b, \rho) \) that encrypts \( \mathrm{q} \in \mathcal{R}_{q,r} \) via the secret key $\mathbf{s} = (s_u)_{0 \leq u < s} \in \mathcal{R}_{q,r}^s$. Here $b \in \mathcal{R}_{q,r}$ and $\rho$ is a $128$-bit seed string. $\mathtt{swk}_j = (\mathtt{swk}_{j,u})_{0 \leq u < s} \in (\mathcal{R}_{qp,d}^2)^s$ are the RLWE switching keys where $\mathtt{swk}_{j,u}$ switches from $\tilde{s_u}$ to $\varphi_j^{-1}(s')$ where $s' \in \mathcal{R}_{*,d}$ is the target RLWE secret key. Here $\mathtt{GenA}$ generates the $a$-part of the MLWE ciphertext from the $128$-bit seed $\rho$, and $\mathtt{ModUp}$ and $\mathtt{ModDown}$ are the typical homomorphic base conversions from $q$ to $qp$ and from $qp$ to $q$.
    \ENSURE RLWE ciphertexts $(ct_j)_{0 \leq j < r}$ that encrypt \( \left( \varphi_{j}(r^{-1} \cdot \mathrm{q}) \right)_{0 \leq j < r} \), i.e. polynomial of degree \( d \) in \( \mathcal{R}_q \) with \( X^{2j+1} \) automorphism operations for \( 0\leq j < r \).
    \STATE \( \mathbf{a} = (a_u)_{0 \leq u < s} \in \mathcal{R}_{q,r}^s \gets \mathtt{GenA}(\rho) \)
    \STATE \( (b, \mathbf{a}) \gets r^{-1} \cdot (b, \mathbf{a}) \) mod \( q \)
    \STATE \( \mathbf{\hat{a}} = (\hat{a}_u)_{0 \leq u < s} \in \mathcal{R}_{qp,r}^s \gets \left(\mathtt{ModUp}(a_u) \right)_{0 \leq u < s} \)
    \FOR{$j=0$ to $r-1$}
        \STATE \( ct_{j} \in \mathcal{R}_{qp,d}^2 \gets \sum_{u=0}^{s-1}(\hat{a}_i \cdot \mathtt{swk}_{j,u}) \)
        \STATE \( ct_{j} \gets \mathtt{ModDown}(ct_{j}) \)
        \STATE \( ct_{j} \gets \varphi_{j}(ct_{j} + (\tilde{b} \in \mathcal{R}_{q,d},\ 0)) \)
    \ENDFOR
    \RETURN \( ( ct_{j})_{0 \leq j < r} \)
\end{algorithmic}
\end{algorithm}

\begin{algorithm}[H]
\caption{\texttt{Cache}}
\label{alg:cache}
\begin{algorithmic}[1]
    \REQUIRE Key (seeded) MLWE ciphertexts \( ( b_{i}, \rho_{i} ) \) that encrypts \( k_{i} \in \mathcal{R}_{q,r} \) via the secret key $\mathbf{s} = (s_u)_{0 \leq u < s} \in \mathcal{R}_{q,r}^s$, for each $0 \leq i < d$. Here $b_{i} \in \mathcal{R}_{q,r}$ and $\rho_{i}$ is a $128$-bit seed string. $\mathtt{swk}_j = (\mathtt{swk}_{j,u})_{0 \leq u < s} \in (\mathcal{R}_{qp,d}^2)^s$ are the RLWE switching keys where $\mathtt{swk}_{j,u}$ switches from $\varphi_{j}(\tilde{s_i})$ to $s'$ where $s' \in \mathcal{R}_{*,d}$ is the target RLWE secret key. Here $\mathtt{GenA}$ generates the $a$-part of the MLWE ciphertext from the $128$-bit seed $\rho$, and $\mathtt{ModUp}$ and $\mathtt{ModDown}$ are the typical homomorphic base conversions from $q$ to $qp$ and vice versa, respectively. Let $\mathbf{B} \in \mathcal{R}_{q,d}^{r \times r}$ be the matrix as defined in Equation~\ref{eq:defB}. 
    \ENSURE RLWE ciphertexts $(ct_j''')_{0 \leq j < r} \in (\mathcal{R}_{q,d}^{2})^r $ that encrypt \( \left(\sum_{i=0}^{d-1} \varphi_{j}(\tilde{k}_{i})X^{i} \right)_{0 \leq j < r} \).
    \FOR{$i=0$ to $d-1$}
        \STATE \( \mathbf{a}_{i} = (a_{i,u})_{0 \leq u < s} \in \mathcal{R}_{q,r}^s \gets \mathtt{GenA}(\rho_{i})  \)
        \STATE \( ( b_{i}, \mathbf{a}_{i} ) \gets r^{-1} \cdot ( b_{i}, \mathbf{a}_{i} ) \) mod \( q \)
    \ENDFOR
    \FOR{$j=0$ to $r-1$}
        \STATE 
        \(
        ( b_{j}', \mathbf{a}_{j}' ) \in \mathcal{R}_{q,d}^{s+1} \gets \left( \sum_{v=0}^{s-1}\tilde{b}_{v+sj}\cdot X^{v}, \left( \sum_{v=0}^{s-1} \tilde{a}_{(v + sj),u} \cdot X^v \right)_{0 \leq u < s} \right)
        \)
    \ENDFOR
    \STATE \( \mathbf{ct}' \in (\mathcal{R}_{q,d}^{s+1})^r \gets ( b_{j}', \mathbf{a}_{j}' )_{0 \leq j < r} \)
    \STATE \( \mathbf{ct'} \in (\mathcal{R}_{q,d}^{s+1})^r \gets \mathbf{B} \cdot \mathbf{ct'} \)
    \FOR{$j=0$ to $r-1$}
        \STATE \( ct_{j}'' = (b_j'', \mathbf{a}_j'') \in \mathcal{R}_{q,d} \times \mathcal{R}_{q,d}^s \gets \varphi_{j,r}\left(\mathbf{ct}'[j]\right) \)
        \STATE \( \mathbf{\hat{a}}_{j}'' = (\hat{a}_{j,u}'')_{0 \leq u < s} \in \mathcal{R}_{qp,d}^s \gets \mathtt{ModUp}(\mathbf{a}_{j}'') \)
        \STATE \( ct_{j}''' \in \mathcal{R}_{qp,d}^2 \gets \sum_{u=0}^{s-1}(\hat{a}_{j,u}'' \cdot \mathtt{swk}_{j,u}) \)
        \STATE \( ct_{j}''' \in \mathcal{R}_{q,d}^2 \gets \mathtt{ModDown}(ct_{j}''') \)
        \STATE \( ct_{j}''' \gets ct_{j}''' + ( b_{j}'' \in \mathcal{R}_{q,d},\ 0) \)
        \STATE \( ct_{j}''' \gets r \cdot ct_{j}''' \) mod \( q \)
    \ENDFOR
    \RETURN \( ( ct_{j}''')_{0 \leq j < r} \)
\end{algorithmic}
\end{algorithm}

\begin{algorithm}[H]
\caption{\texttt{Score}}
\label{alg:score}
\begin{algorithmic}[1]
    \REQUIRE \texttt{Decompose}d query ciphertexts $\mathbf{ct}_q \in (\mathcal{R}_{q,d}^2)^r$, \texttt{Cache}d key ciphertexts $\mathbf{ct}_k \in (\mathcal{R}_{q,d} ^2)^r$.
    \ENSURE A RLWE ciphertext $ct_{out}$ encrypting the resulting score polynomial $\sum_{j=0}^{d-1} \sigma_j X^j$.
    \STATE $ct_{out} \gets \mathtt{Relin}(\sum_{i=0}^{r-1} \mathbf{ct}_q[i] \otimes \mathbf{ct}_k[i])$ 
    \RETURN $ct_{out}$
\end{algorithmic}
\end{algorithm}

\subsection{Private Information Retrieval}

We extend our Secure Inner Product method to support Private Information Retrieval (PIR). Similar to SPIRAL~\cite{spiralpir}, we treat the database as a matrix. The protocol requires the client to send two encrypted queries: one selecting the target row and the other selecting the target column, each containing a one hot vector at the corresponding index. The server then performs PIR through two sequential applications of the Secure Inner Product protocol. 
However, naively applying the Secure Inner Product protocol in this PIR context introduces a cache invalidation issue. Specifically, while the standalone Secure Inner Product scenario only requires refreshing the cache corresponding to the updated index, PIR necessitates refreshing the entire cache whenever the database changes. This occurs because the output from the first stage acts as the key for the second stage. To address this, we modify our protocol by applying the inverse butterfly operation—originally intended for use on the key—to the decomposed query instead.

In our experimental setting using a Fast network (see Section~\ref{sec:resources}), the modified PIR protocol achieves an end-to-end retrieval latency of under 700 ms for databases consisting of $2^{20}$ records, each sized at 1 KiB. Consequently, we demonstrate that our approach efficiently supports a secure vector database of 1 GiB containing 1 million records with 96 dimensions each, achieving an end-to-end latency below 1 second.

%% file: sections/B_experimental_setup.tex
\section{Experimental Setup}
\label{sec:appendix_experimental_setup}

\niloofar{one other thing that needs to be threaded carefully is the dataset mismatch between 4.1 and 4.2. The fact that LAION is used only in section 4.2 may confuse ppl. We need to justify/explain that.}

\subsection{Socratic Chain-of-Thought Reasoning}

We empirically evaluate the effectiveness of our reasoning framework in addressing the computational limitations of local models.
Experiments are conducted on two QA-focused benchmarks:
LoCoMo, which simulates personal assistant scenarios, and MediQ, which simulates medical consultation scenarios.
Both tasks require retrieving relevant private user data and performing complex reasoning to arrive at a final answer.
We compare our framework against two categories of baselines:
Golden Baselines assume no privacy constraints, allowing private data to be directly passed to remote models.
We use GPT-4o (R1), Gemini-1.5-Pro (R2), and Claude-3.5-Sonnet (R3), which cannot be run locally but offer strong reasoning capabilities.
Local-only Baselines assume strong privacy constraints, requiring the entire inference process to be carried out by local models.
We use Llama-3.2-1B (L1), Llama-3.2-3B (L2), and Llama-3.1-8B (L3), which are lightweight enough for local execution but less capable in complex reasoning tasks.
The goal of our reasoning framework is to improve the performance of local-only baselines by leveraging model collaboration and delegated reasoning, aiming to approach the performance of the golden baselines.

\subsection{Homomorphically Encrypted Vector Database}

We examine whether vector search can be performed accurately and efficiently over encrypted data using homomorphic encryption. Our goal is to match the quality and latency of plaintext vector search while ensuring that both queries and database contents remain private. The encrypted vector database is implemented using HEXL~\cite{hexl} and evaluated in  in the same Google Cloud Platform configuration used by Compass~\cite{compass} for a fair comparison: an n2-standard-8 instance (8 vCPUs @ 2.8 GHz, 32 GB RAM) as the client and an n2-highmem-64 instance (64 vCPUs @ 2.8 GHz, 512 GB RAM) as the server, co-located in the same region/zone. Using Linux Traffic Control, we emulate two network regimes: Fast (3 Gbps, 1 ms Round Trip Time (RTT)) and Slow (400 Mbps, 80 ms RTT) to isolate the impact of bandwidth and latency. We use 10k query vectors and 1M key vectors from Deep1B (96D) and LAION (512D), as well as the entire LoCoMo dataset (768D). For search accuracy, we report mean/max inner product error, MRR@10, and 1-Recall@k. For latency, we measure end-to-end CPU runtime. All speed measurements assume that both the query and the keys are ciphertexts and employ parameters that satisfy IND-CPA 128-bit security. To evaluate storage, we analyze ciphertext overhead and apply packing optimizations.

\subsection{Hyperparameter Selection}

To evaluate Socratic Chain-of-Thought Reasoning, we set the temperature of all language models to zero to ensure reproducibility.
We use top-k retrieval with reranking based on vector similarity scores. 
We set \( k \) to 5 for LoCoMo and 20 for MediQ, as the maximum number of ground truth retrievals varies across datasets.

\subsection{Model Selection}

We employ DRAGON~\cite{dragon} as the retriever because it outperforms other candidates, such as DPR~\cite{dpr}, Contriever~\cite{contriever}, and Instructor~\cite{instructor}, on our chosen datasets.  
It represents data as 768-dimensional vectors, and the inner product between two vectors is used to compute the similarity score.  
For the remote models, we use GPT-4o (R1)~\cite{gpt4o}, Gemini-1.5-Pro (R2)~\cite{gemini}, and Claude-3.5-Sonnet (R3)~\cite{claude}, representing the most powerful closed API language models currently available. These models are assumed to run in a public cloud environment.
For the local models, we select Llama-3.2-1B (L1), Llama-3.2-3B (L2), and Llama-3.1-8B (L3)~\cite{llama3}, which are lightweight enough to be deployed on edge devices.
These models reflect realistic constraints for privacy-preserving, on-device inference.
This selection enables a clear evaluation of our framework, balancing reasoning capability with privacy constraints.

\subsection{Benchmark Selection}

We report the performance of Socratic Chain-of-Thought Reasoning on two benchmarks.  
The first, LoCoMo~\cite{locomo}, is a benchmark designed to test language models in long-term dialogues.  
It simulates an everyday personal assistance scenario, where personal information is gradually accumulated in a vector database through extended observation.  
On LoCoMo, we evaluate (1) the remote models's impact on retrieval using Recall@5 and (2) its enhancement of response quality through improved response generation, measured by the F1 score.  
We use only the single-hop QA and multi-hop QA datasets out of the total five datasets in LoCoMo, as these are the only datasets suitable for our scenario.  
The second benchmark, MediQ~\cite{mediq}, presents a more specialized scenario focused on medical consultation, where privacy risks are directly at odds with the need for access to a patient's personal context.
MediQ is a multiple-choice question-answering dataset, so we evaluate generation accuracy using the exact match metric.
Since MediQ lacks retrieval annotations, we do not report retrieval metric for this benchmark.

We report the performance of the homomorphically encrypted vector database on standard retrieval benchmarks.
To assess the scalability of encrypted storage and search, we selected a sufficiently large dataset.
We used the top 10k query vectors and 1M key vectors from Deep1B~\cite{deep1b} and LAION~\cite{laion}, represented as 96-dimensional and 512-dimensional vectors respectively.
For LoCoMo~\cite{locomo}, we used the entire dataset, which consists of 1,742 query vectors and 4,972 key vectors, each represented as a 768-dimensional vector.

\subsection{Metric Selection}

For the Socratic Chain-of-Thought Reasoning, we focus on measuring the quality of the generated answers. On the LoCoMo benchmark, we report the F1 score, which captures token-level overlap between generated and ground-truth responses in long-context dialogues. On the MediQ benchmark, we report exact match accuracy, as the task involves multiple-choice question answering and requires strict correctness. These metrics enable us to quantify the impact of delegating complex reasoning to powerful remote models while keeping sensitive data within a trusted zone.

For the homomorphically encrypted vector database, we evaluate both search accuracy and latency. To assess search accuracy, we compute the mean error and maximum error between the inner product similarity scores produced by encrypted and plaintext searches. Additionally, we report 1-Recall@1 and 1-Recall@5, which represent the proportion of queries for which the top-1 result from the plaintext database is not recovered in the top-1 or top-5 encrypted results. Lower values for these metrics indicate higher retrieval consistency under encryption. To evaluate latency, we measure the average response time of encrypted search queries. All metrics are reported separately for plaintext and ciphertext queries.

%% file: sections/C_compute_resources.tex
\section{Compute Resources}
\label{sec:resources}
For Socratic Chain-of-Thought Reasoning, all experiments were conducted using a single NVIDIA A100 GPU.
Language models from the Llama family were accessed via the Fireworks API~\cite{fireworks}, while other closed API models, including those from OpenAI, Gemini, and Claude, were accessed through their respective APIs.
Our homomorphically encrypted vector database was implemented using HEXL~\cite{hexl} and evaluated under the same Google Cloud Platform configuration used by Compass~\cite{compass} to ensure a fair comparison: an n2-standard-8 instance (8 vCPUs @ 2.8 GHz, 32 GB RAM) was used as the client, and an n2-highmem-64 instance (64 vCPUs @ 2.8 GHz, 512 GB RAM) was used as the server, both co-located in the same region and zone.
To emulate realistic networking conditions, we used Linux Traffic Control to simulate two environments: \textbf{Fast} (3 Gbps bandwidth, 1 ms round-trip time and \textbf{Slow} (400 Mbps bandwidth, 80 ms round-trip time).
The following commands were used to apply these network configurations to the server.

\paragraph{Fast Network}
\begin{verbatim}
tc qdisc add dev ens4 root netem delay 1ms
tc qdisc add dev ens4 root handle 1: htb default 30
tc class add dev ens4 parent 1: classid 1:1 htb rate 3096mbps
tc class add dev ens4 parent 1: classid 1:2 htb rate 3096mbps
tc filter add dev ens4 protocol ip parent 1:0 prio 1 u32 \
match ip dst $CLIENT_IP flowid 1:1
tc filter add dev ens4 protocol ip parent 1:0 prio 1 u32 \
match ip src $CLIENT_IP flowid 1:2
\end{verbatim}

\paragraph{Slow Network}
\begin{verbatim}
tc qdisc add dev ens4 root netem delay 80ms
tc qdisc add dev ens4 root handle 1: htb default 30
tc class add dev ens4 parent 1: classid 1:1 htb rate 400mbps
tc class add dev ens4 parent 1: classid 1:2 htb rate 400mbps
tc filter add dev ens4 protocol ip parent 1:0 prio 1 u32 \
match ip dst $CLIENT_IP flowid 1:1
tc filter add dev ens4 protocol ip parent 1:0 prio 1 u32 \
match ip src $CLIENT_IP flowid 1:2
\end{verbatim}

%% file: sections/D_qualitative_analysis.tex
\section{Qualitative Analysis}
\label{sec:qualitative_analysis}

We present qualitative examples from the LoCoMo and MediQ benchmarks to illustrate how our system improves response quality under strict privacy constraints. By delegating sub-query generation and chain-of-thought reasoning to a powerful remote model, and executing final response generation locally, our framework ensures that sensitive data never leaves the trusted zone while still benefiting from advanced reasoning capabilities.

\subsection{LoCoMo}

\textbf{User Query.} \textit{“What motivated Caroline to pursue counseling?”}

This query requires linking the user’s past personal experiences to her career decisions, as this information is often buried in long conversational histories.

\textbf{Sub-Query Generation by Remote Model.}
The remote model generated sub-queries such as:
\textit{“Has Caroline discussed any impactful personal experiences related to her career?”}
\textit{“Did she mention an interest in counseling in past conversations?”}

These sub-queries were embedded on the local client and used to search the homomorphically encrypted vector database.

\textbf{Encrypted Search from Private Records.}
The search retrieved a key statement:
\textit{“My own journey and the support I got made a huge difference... I saw how counseling and support groups improved my life.”}

\textbf{Chain-of-Thought Reasoning from Remote Model.}
The model suggested this reasoning guideline:
\textit{“When personal growth or transformation is attributed to support or counseling, infer a connection between that experience and a career motivation to help others.”}

\textbf{Response Generation by Local Model.}
Using the retrieved memory and the reasoning instruction, the local model generated the following answer:
\textit{“Caroline was motivated to pursue counseling because of her own journey and the support she received, particularly through counseling and support groups.”}

\subsection{MediQ}

\textbf{User Query.} \textit{“I’ve been feeling more forgetful lately and have started falling more often. What should I do?”}

This query suggests a combination of cognitive and physical decline, potentially indicating an underlying neurological issue. Proper assessment requires integration of personal medical context and symptom history.

\textbf{Sub-Query Generation by Remote Model.}
The remote model generated targeted follow-up questions, including:
\textit{“Is there any record of short-term memory impairment?”}
\textit{“Have the falls become more frequent or severe over time?”}
\textit{“Are there other neurological symptoms noted in the history?”}

\textbf{Encrypted Search from Private Records.}
These sub-queries were executed on encrypted medical records, retrieving relevant notes such as:
\textit{“I couldn’t remember any of the five things the doctor asked me to recall after ten minutes.”}
\textit{“I’ve been falling more often lately, and it feels like it’s getting worse.”}

\textbf{Chain-of-Thought Reasoning from Remote Model.}
The remote model provided the following reasoning instruction to the local model:
\textit{“When both progressive memory loss and increased frequency of falls are reported, evaluate for possible neurodegenerative conditions and recommend medical assessment.”}

\textbf{Response Generation by Local Model.}
Based on the retrieved data and reasoning instruction, the local model generated the following concise response:
\textit{“Parkinson’s disease.”}

These examples demonstrate that our framework enables local models to generate informed, context-sensitive responses by leveraging powerful remote models for high-level reasoning. Throughout the process, sensitive user data remains local, ensuring strong privacy guarantees while maintaining or even improving response quality.

%% file: sections/E_prompt_templates.tex
\section{Prompt Templates}

\newcommand{\promptSubstitution}[1]{\textcolor{red}{{#1}}}
\newcommand{\promptBox}[1]{\fbox{\parbox{0.9\linewidth}{\texttt{#1}}}}
\setlength{\fboxsep}{0.5em}

For sub-query generation in both the baselines and Socratic Chain-of-Thought Reasoning, we used the prompt shown in Figure~\ref{fig:prompt_sub_query_generation}.  
For response generation in the baselines, the prompt in Figure~\ref{fig:prompt_response_generation} was used.  
For Socratic Chain-of-Thought Reasoning, chain-of-thought generation was performed using the prompt in Figure~\ref{fig:prompt_cot_generation}, and response generation used the prompt in Figure~\ref{fig:prompt_cot_execution}.
The prompts include substitution keys, which are described in Table~\ref{tab:prompt_substitutions}.

\input{tables/E_prompt_substitutions}
\clearpage
\input{figures/prompts/sub_query_generation}
\clearpage
\input{figures/prompts/response_generation}
\input{figures/prompts/cot_generation}
\clearpage
\input{figures/prompts/cot_execution}
\clearpage

\section{Additional MediQ Analysis}
\label{sec:appendix_additional_mediq_analysis}
As shown in Table~\ref{tab:4_experiment_ablation_1}, the Remote-Only Baseline with Socratic Chain-of-Thought Reasoning performs worse than the standard Remote-Only Baseline on MediQ.
To understand the cause of this drop, we conducted a detailed qualitative analysis of the model's inputs and outputs.
As a result, we found that R1 (GPT-4o), when generating chain-of-thought reasoning, often included the most likely answer without considering the user's personal context.
As a result, L1 (Llama-3.2-1B) became strongly biased toward this uncontextualized answer and also ignored the user's personal context.
To address this issue, we added explicit rules to the prompt—shown in Figure~\ref{fig:prompt_cot_generation_mediq}—to reduce this bias and re-ran the experiment under this setup only.
With this adjustment, performance improved from 67.3 to 77.0, indicating that the bias was partially mitigated.
\input{figures/prompts/cot_generation_mediq}

%% file: tables/E_prompt_substitutions.tex
\begin{table}[h]
    \centering
    \begin{tabular}{c | p{0.3\textwidth} p{0.4\textwidth}}
        \toprule
        \textbf{Key} & \textbf{Description} & \textbf{Illustrative Example}  \\
        \midrule
        \midrule
        \{user\_input\} & User input & \texttt{I have a fever and a cough. What disease do I have?} \\
        \midrule
        \{options\} & Multiple-choice option. Formatted as bulleted list. For open ended questions, this is replaced with \texttt{Empty} instead. & \texttt{- Common cold \newline
        - Flu \newline
        - Strep throat} \\
        \midrule
        \{personal\_context\} & List of retrieved personal contexts in descending order of importance, one item on each line. & \texttt{In January 30th, user consumed a half gallon of ice cream. \newline User enjoys cold drink, even in winter. \newline User spends most of the time in their place alone.} \\
        \midrule
        \{personal\_context\_json\} & List of retrieved personal contexts in descending order of importance, as JSON-formatted array of strings. & \texttt{[\newline \null\qquad "In January 30th, user consumed a half gallon of ice cream.",\newline \null\qquad "User enjoys cold drink, even in winter.",\newline \null\qquad "User spends most of the time in their place alone."\newline]} \\
        \midrule
        \{generated\_reasoning\} & The output of reasoning generation step. & (omitted) \\
        \bottomrule
    \end{tabular}
    \vspace{1.0em}
    \caption{Substitutions for our prompts. Whenever the listed substitution keys appear on our prompt template, they are substituted into the actual values as described on the right side of the table.}
    \label{tab:prompt_substitutions}
\end{table}

%% file: figures/prompts/sub_query_generation.tex
\begin{figure}
    \centering
    \promptBox{You are a sub-query generator.\\
\\
1. You are given a query and a list of possible options.\\
2. Your task is to generate 3 to 5 sub-queries that help retrieve personal context relevant to answering the query.\\
3. Each sub-query should be answerable based on the user\textquotesingle{}s personal context.\\
4. Ensure the sub-queries cover different aspects or angles of the query.\\
5. If the options text says \textquotesingle{}Empty,\textquotesingle{} it means no options are provided.\\
\\
Please output the sub-queries one sub-query each line, in the following format:\\
"Sub-query 1 here"\\
"Sub-query 2 here"\\
"Sub-query 3 here"\\
\\
Example 1)\\
\\
\#\# Query\\
I have a fever and a cough. What disease do I have?\\
\\
\#\# Options\\
Common cold\\
Flu\\
Strep throat\\
\\
\#\#\# Sub-queries\\
"Have user visited any countries in Africa recently?"\\
"Have user eat any cold food recently?"\\
"Have user been in contact with anyone who has a COVID-19 recently?"\\
\\
Test Input)\\
\\
\#\#\# Query\\
\promptSubstitution{\{user\_input\}}\\
\\
\#\#\# Options\\
\promptSubstitution{\{options\}}\\
\\
\#\#\# Sub-queries}
    \caption{Prompt used for sub-query generation in both the baselines and the socratic chain-of-thought reasoning.}
    \label{fig:prompt_sub_query_generation}
\end{figure}

%% file: figures/prompts/response_generation.tex
\begin{figure}
    \centering
    \promptBox{You are a question answering model.\\
\\
1. You are given a personal context, a query, and a list of possible options.\\
2. Your task is to generate an answer to the query based on the user's personal context.\\
3. You should generate an answer to the query by referring to the personal context where relevant.\\
4. If the options text says \textquotesingle{}Empty,\textquotesingle{} it means no options are provided.\\
5. If the options are not empty, simply output one of the answers listed in the options without any additional explanation.\\
6. Never output any other explanation. Just output the answer.\\
7. If option follows a format like \textquotesingle{}[A] something\textquotesingle{}, then output something as the answer instead of A.\\
\\
Test Input)\\
\\
\\
\#\#\# Personal Context\\
\promptSubstitution{\{personal\_context\}}\\
\\
\#\#\# Question\\
\promptSubstitution{\{user\_input\}}\\
\\
\#\#\# Options\\
\promptSubstitution{\{options\}}\\
\\
\#\#\# Answer
}
    \caption{Prompt used for response generation in the baselines.}
    \label{fig:prompt_response_generation}
\end{figure}

%% file: figures/prompts/cot_generation.tex
\begin{figure}
    \centering
    \promptBox{Your task is to provide good reasoning guide for students.\\
\\
You are a chain-of-thought generator.\\
1. You are given a query and a list of possible options.\\
2. Your task is to provide a step-by-step reasoning guide to help a student answer the query.\\
3. The reasoning guide should clearly show your reasoning process so that the student can easily apply it to their query.\\
4. Analyze the query and write a reasoning guide for the student to follow.\\
5. If there is a lack of information relevant to the query, you must identify the missing elements as "VARIABLES" and write the guide on a case-by-case basis.\\
6. If the options text says \textquotesingle{}Empty,\textquotesingle{} it means no options are provided.\\
\\
Test Input)\\
\\
\#\#\# Query\\
\promptSubstitution{\{user\_input\}}\\
\\
\#\#\# Options\\
\promptSubstitution{\{options\}}\\
\\
\#\#\# Chain-of-Thought}
    \caption{Prompt used for chain-of-thought generation in the socratic chain-of-thought reasoning.}
    \label{fig:prompt_cot_generation}
\end{figure}

%% file: figures/prompts/cot_execution.tex
\begin{figure}
    \centering
    \promptBox{You are a question answering model.\\
\\
1. Your task is to answer the query based on the teacher\textquotesingle{}s chain-of-thought decision guide, using additional personal context.\\
2. Read the chain-of-thought decision guide carefully.\\
3. If the decision guide contains "VARIABLES" that may affect the outcome, extract them and determine their values based on the personal context.\\
4. Then, follow the decision guide and apply the extracted variables appropriately to derive the final answer.\\
5. The final answer must be preceded by \textquotesingle{}\#\#\# Answer\textquotesingle{}, and your response must end immediately after the answer.\\
6. If the options text says \textquotesingle{}Empty,\textquotesingle{} it means no options are provided.\\
7. If the options are not empty, simply output one of the answers listed in the options without any additional explanation.\\
8. Never output any other explanation. Just output the answer.\\
9. If option follows a format like \textquotesingle{}[A] something\textquotesingle{}, then output something as the answer instead of A.\\
\\
\#\#\# Personal Context\\
\{personal\_context\_json\}\\
\\
\#\#\# Chain-of-Thought\\
\{cot\}\\
\\
\#\#\# Query\\
\promptSubstitution{\{user\_input\}}\\
\\
\#\#\# Options\\
\promptSubstitution{\{options\}}\\
\\
\#\#\# Answer}
    \caption{Prompt used for response generation in the socratic chain-of-thought reasoning.}
    \label{fig:prompt_cot_execution}
\end{figure}

%% file: figures/prompts/cot_generation_mediq.tex
\begin{figure}[h]
    \centering
    \promptBox{Your task is to provide good reasoning guide for students.\\
\\
You are a chain-of-thought generator.\\
1. You are given a query and a list of possible options.\\
2. Your task is to provide a step-by-step reasoning guide to help a student answer the query.\\
3. The reasoning guide should clearly show your reasoning process so that the student can easily apply it to their query.\\
4. Analyze the query and write a reasoning guide for the student to follow.\\
5. The student may have less domain knowledge than you, but they have more context about the situation.\\
6. If there is a lack of information relevant to the query, you must identify the missing elements as "VARIABLES" and write the guide on a case-by-case basis.\\
7. Since you don’t have full context about the situation, your goal is not to choose a final answer but to present a set of possible answers along with the reasoning steps that could lead to each one.\\
8. If the options text says \textquotesingle{}Empty,\textquotesingle{} it means no options are provided.\\
\\
Test Input)\\
\\
\#\#\# Query\\
\promptSubstitution{\{user\_input\}}\\
\\
\#\#\# Options\\
\promptSubstitution{\{options\}}\\
\\
\#\#\# Chain-of-Thought}
\caption{Prompt used for chain-of-thought generation in the additional MediQ analysis.}
\label{fig:prompt_cot_generation_mediq}
\end{figure}

%% file: root.bbl
\begin{thebibliography}{87}
\providecommand{\natexlab}[1]{#1}
\providecommand{\url}[1]{\texttt{#1}}
\expandafter\ifx\csname urlstyle\endcsname\relax
  \providecommand{\doi}[1]{doi: #1}\else
  \providecommand{\doi}{doi: \begingroup \urlstyle{rm}\Url}\fi

\bibitem[Abadi et~al.(2016)Abadi, Chu, Goodfellow, McMahan, Mironov, Talwar, and Zhang]{Abadi2016}
Martín Abadi, Andy Chu, Ian~J. Goodfellow, H.~Brendan McMahan, Ilya Mironov, Kunal Talwar, and Li~Zhang.
\newblock Deep learning with differential privacy.
\newblock In \emph{Proceedings of the 2016 {ACM} {SIGSAC} Conference on Computer and Communications Security (CCS)}, pages 308--318, 2016.
\newblock \doi{10.1145/2976749.2978318}.

\bibitem[Anthropic(2024)]{claude}
Anthropic.
\newblock Claude 3.5 sonnet, 2024.
\newblock URL \url{https://www.anthropic.com/news/claude-3-5-sonnet}.
\newblock Accessed: 2025-01-29.

\bibitem[Asi et~al.(2024)Asi, Boemer, Genise, Mughees, Ogilvie, Rishi, Rothblum, Talwar, Tarbe, Zhu, et~al.]{wally}
Hilal Asi, Fabian Boemer, Nicholas Genise, Muhammad~Haris Mughees, Tabitha Ogilvie, Rehan Rishi, Guy~N Rothblum, Kunal Talwar, Karl Tarbe, Ruiyu Zhu, et~al.
\newblock Scalable private search with wally.
\newblock \emph{arXiv preprint arXiv:2406.06761}, 2024.

\bibitem[Babenko and Lempitsky(2016)]{deep1b}
Artem Babenko and Victor Lempitsky.
\newblock Efficient indexing of billion-scale datasets of deep descriptors.
\newblock In \emph{Proceedings of the IEEE Conference on Computer Vision and Pattern Recognition}, 2016.

\bibitem[Bae et~al.(2023)Bae, Cheon, Kim, Park, and Stehl{\'e}]{hermes}
Youngjin Bae, Jung~Hee Cheon, Jaehyung Kim, Jai~Hyun Park, and Damien Stehl{\'e}.
\newblock Hermes: Efficient ring packing using mlwe ciphertexts and application to transciphering.
\newblock In Helena Handschuh and Anna Lysyanskaya, editors, \emph{Advances in Cryptology -- CRYPTO 2023}, pages 37--69, Cham, 2023. Springer Nature Switzerland.
\newblock ISBN 978-3-031-38551-3.

\bibitem[Bianchi et~al.(2023)Bianchi, Suzgun, Attanasio, Röttger, Jurafsky, Hashimoto, and Zou]{Bianchi2023}
Federico Bianchi, Mirac Suzgun, Giuseppe Attanasio, Paul Röttger, Dan Jurafsky, Tatsunori Hashimoto, and James Zou.
\newblock Safety-tuned llamas: Lessons from improving the safety of large language models that follow instructions.
\newblock \emph{arXiv preprint arXiv:2309.07875}, 2023.
\newblock URL \url{https://arxiv.org/abs/2309.07875}.

\bibitem[Boemer et~al.(2021)Boemer, Kim, Seifu, DM~de Souza, and Gopal]{hexl}
Fabian Boemer, Sejun Kim, Gelila Seifu, Fillipe DM~de Souza, and Vinodh Gopal.
\newblock Intel hexl: accelerating homomorphic encryption with intel avx512-ifma52.
\newblock In \emph{Proceedings of the 9th on Workshop on Encrypted Computing \& Applied Homomorphic Cryptography}, 2021.

\bibitem[Bonnetain et~al.(2019)Bonnetain, Naya-Plasencia, and Schrottenloher]{bonnetain2019quantum}
Xavier Bonnetain, Mar{\'\i}a Naya-Plasencia, and Andr{\'e} Schrottenloher.
\newblock Quantum security analysis of aes.
\newblock \emph{IACR Transactions on Symmetric Cryptology}, 2019\penalty0 (2):\penalty0 55--93, 2019.

\bibitem[Bos et~al.(2018)Bos, Ducas, Kiltz, Lepoint, Lyubashevsky, Schanck, Schwabe, Seiler, and Stehl{\'e}]{kyber}
Joppe Bos, L{\'e}o Ducas, Eike Kiltz, Tancr{\`e}de Lepoint, Vadim Lyubashevsky, John~M Schanck, Peter Schwabe, Gregor Seiler, and Damien Stehl{\'e}.
\newblock Crystals-kyber: a cca-secure module-lattice-based kem.
\newblock In \emph{2018 IEEE European Symposium on Security and Privacy (EuroS\&P)}, pages 353--367. IEEE, 2018.

\bibitem[Bossuat et~al.(2021)Bossuat, Mouchet, Troncoso-Pastoriza, and Hubaux]{hoist}
Jean-Philippe Bossuat, Christian Mouchet, Juan Troncoso-Pastoriza, and Jean-Pierre Hubaux.
\newblock Efficient bootstrapping for approximate homomorphic encryption with non-sparse keys.
\newblock In \emph{Annual International Conference on the Theory and Applications of Cryptographic Techniques}, pages 587--617. Springer, 2021.

\bibitem[Bossuat et~al.(2024)Bossuat, Cammarota, Chillotti, Curtis, Dai, Gong, Hales, Kim, Kumara, Lee, Lu, Maple, Pedrouzo-Ulloa, Player, Polyakov, Lopez, Song, and Yhee]{security-guidelines}
Jean-Philippe Bossuat, Rosario Cammarota, Ilaria Chillotti, Benjamin~R. Curtis, Wei Dai, Huijing Gong, Erin Hales, Duhyeong Kim, Bryan Kumara, Changmin Lee, Xianhui Lu, Carsten Maple, Alberto Pedrouzo-Ulloa, Rachel Player, Yuriy Polyakov, Luis Antonio~Ruiz Lopez, Yongsoo Song, and Donggeon Yhee.
\newblock Security guidelines for implementing homomorphic encryption.
\newblock Cryptology {ePrint} Archive, Paper 2024/463, 2024.
\newblock URL \url{https://eprint.iacr.org/2024/463}.

\bibitem[Brakerski et~al.(2014)Brakerski, Gentry, and Vaikuntanathan]{brakerski2014leveled}
Zvika Brakerski, Craig Gentry, and Vinod Vaikuntanathan.
\newblock (leveled) fully homomorphic encryption without bootstrapping.
\newblock \emph{ACM Transactions on Computation Theory}, 6\penalty0 (3):\penalty0 1--36, 2014.

\bibitem[Cappelli et~al.(2012)Cappelli, Moore, and Trzeciak]{cappelli2012cert}
Dawn~M Cappelli, Andrew~P Moore, and Randall~F Trzeciak.
\newblock \emph{The CERT guide to insider threats: how to prevent, detect, and respond to information technology crimes}.
\newblock Addison-Wesley, 2012.

\bibitem[Chen et~al.(2021)Chen, Dai, Kim, and Song]{cdks}
Hao Chen, Wei Dai, Miran Kim, and Yongsoo Song.
\newblock Efficient homomorphic conversion between (ring) lwe ciphertexts.
\newblock In \emph{International conference on applied cryptography and network security}, pages 460--479. Springer, 2021.

\bibitem[Chen et~al.(2023)Chen, Zaharia, and Zou]{Chen2023Frugal}
Lingjiao Chen, Matei Zaharia, and James Zou.
\newblock Frugalgpt: How to use large language models while reducing cost and improving performance.
\newblock \emph{arXiv preprint arXiv:2305.05176}, 2023.
\newblock URL \url{https://arxiv.org/abs/2305.05176}.

\bibitem[Chen et~al.(2024)Chen, Zaharia, and Zou]{chen2023frugalgpt}
Lingjiao Chen, Matei Zaharia, and James Zou.
\newblock Frugal{GPT}: How to use large language models while reducing cost and improving performance.
\newblock In \emph{International Conference on Learning Representations (ICLR)}, 2024.

\bibitem[Cheon et~al.(2017)Cheon, Kim, Kim, and Song]{ckks}
Jung~Hee Cheon, Andrey Kim, Miran Kim, and Yongsoo Song.
\newblock Homomorphic encryption for arithmetic of approximate numbers.
\newblock In \emph{Advances in Cryptology--ASIACRYPT 2017: 23rd International Conference on the Theory and Applications of Cryptology and Information Security, Hong Kong, China, December 3-7, 2017, Proceedings, Part I 23}, 2017.

\bibitem[Chowdhury et~al.(2025)Chowdhury, Glukhov, Anshumaan, Chalasani, Papernot, Jha, and Bellare]{chowdhury2025pr}
Amrita~Roy Chowdhury, David Glukhov, Divyam Anshumaan, Prasad Chalasani, Nicolas Papernot, Somesh Jha, and Mihir Bellare.
\newblock Preempt: Sanitizing sensitive prompts for llms.
\newblock \emph{arXiv preprint arXiv:2504.05147}, 2025.

\bibitem[Ding et~al.(2024)Ding, Mallick, Wang, Sim, Mukherjee, Ruhle, Lakshmanan, and Awadallah]{Ding2024}
Dujian Ding, Ankur Mallick, Chi Wang, Robert Sim, Subhabrata Mukherjee, Victor Ruhle, Laks V.~S. Lakshmanan, and Ahmed~Hassan Awadallah.
\newblock Hybrid llm: Cost-efficient and quality-aware query routing.
\newblock \emph{arXiv preprint arXiv:2404.14618}, 2024.
\newblock URL \url{https://arxiv.org/abs/2404.14618}.

\bibitem[Dong et~al.(2023)Dong, Lu, Zheng, Wu, Zhao, Tan, Huang, Hong, Wei, and Chen]{Li2023}
Ye~Dong, Wen{-}jie Lu, Yancheng Zheng, Haoqi Wu, Derun Zhao, Jin Tan, Zhicong Huang, Cheng Hong, Tao Wei, and Wenguang Chen.
\newblock {PUMA}: Secure inference of {LLaMA-7B} in five minutes.
\newblock \emph{CoRR}, abs/2307.12533, 2023.
\newblock \doi{10.48550/arXiv.2307.12533}.

\bibitem[Dou et~al.(2023)Dou, Krsek, Naous, Kabra, Das, Ritter, and Xu]{dou2023reducing}
Yao Dou, Isadora Krsek, Tarek Naous, Anubha Kabra, Sauvik Das, Alan Ritter, and Wei Xu.
\newblock Reducing privacy risks in online self-disclosures with language models.
\newblock \emph{arXiv preprint arXiv:2311.09538}, 2023.

\bibitem[Dubey et~al.(2024)Dubey, Jauhri, Pandey, Kadian, Al-Dahle, Letman, Mathur, Schelten, Yang, Fan, et~al.]{llama3}
Abhimanyu Dubey, Abhinav Jauhri, Abhinav Pandey, Abhishek Kadian, Ahmad Al-Dahle, Aiesha Letman, Akhil Mathur, Alan Schelten, Amy Yang, Angela Fan, et~al.
\newblock The llama 3 herd of models.
\newblock \emph{arXiv preprint arXiv:2407.21783}, 2024.

\bibitem[Engelsma et~al.(2022)Engelsma, Jain, and Boddeti]{hers}
Joshua~J. Engelsma, Anil~K. Jain, and Vishnu~Naresh Boddeti.
\newblock Hers: Homomorphically encrypted representation search.
\newblock \emph{IEEE Transactions on Biometrics, Behavior, and Identity Science}, 4\penalty0 (3):\penalty0 349--360, 2022.
\newblock \doi{10.1109/TBIOM.2021.3139866}.

\bibitem[Gentry(2009{\natexlab{a}})]{Gen09}
Craig Gentry.
\newblock Fully homomorphic encryption using ideal lattices.
\newblock In \emph{Proceedings of the Forty-First Annual ACM Symposium on Theory of Computing}, STOC '09, page 169–178, New York, NY, USA, 2009{\natexlab{a}}. Association for Computing Machinery.
\newblock ISBN 9781605585062.
\newblock \doi{10.1145/1536414.1536440}.
\newblock URL \url{https://doi.org/10.1145/1536414.1536440}.

\bibitem[Gentry(2009{\natexlab{b}})]{gentry2009fully}
Craig Gentry.
\newblock Fully homomorphic encryption using ideal lattices.
\newblock In \emph{Proceedings of the 41st annual ACM symposium on Theory of computing}, pages 169--178, 2009{\natexlab{b}}.

\bibitem[Gilad{-}Bachrach et~al.(2016)Gilad{-}Bachrach, Dowlin, Laine, Lauter, Naehrig, and Wernsing]{GiladBachrach2016}
Ran Gilad{-}Bachrach, Nathan Dowlin, Kim Laine, Kristin Lauter, Michael Naehrig, and John Wernsing.
\newblock Cryptonets: Applying neural networks to encrypted data with high throughput and accuracy.
\newblock In \emph{Proceedings of the 33rd International Conference on Machine Learning (ICML)}, pages 201--210, 2016.

\bibitem[Goldreich(2019)]{goldreich2019foundations}
Oded Goldreich.
\newblock \emph{Foundations of Cryptography: Volume 2, Basic Applications}.
\newblock Cambridge University Press, 2019.

\bibitem[Gupta et~al.(2024)Gupta, Narasimhan, Jitkrittum, Rawat, Menon, and Kumar]{Gupta2024}
Neha Gupta, Harikrishna Narasimhan, Wittawat Jitkrittum, Ankit~Singh Rawat, Aditya~Krishna Menon, and Sanjiv Kumar.
\newblock Language model cascades: Token-level uncertainty and beyond.
\newblock \emph{arXiv preprint arXiv:2404.10136}, 2024.
\newblock URL \url{https://arxiv.org/abs/2404.10136}.

\bibitem[Halevi and Shoup(2018)]{helib-lintrans}
Shai Halevi and Victor Shoup.
\newblock Faster homomorphic linear transformations in helib.
\newblock In \emph{CRYPTO}, 2018.

\bibitem[Hong et~al.(2021)Hong, Kim, Choi, Lee, and Cheon]{kway-sorting}
Seungwan Hong, Seunghong Kim, Jiheon Choi, Younho Lee, and Jung~Hee Cheon.
\newblock Efficient sorting of homomorphic encrypted data with k-way sorting network.
\newblock \emph{IEEE Transactions on Information Forensics and Security}, 16:\penalty0 4389--4404, 2021.
\newblock \doi{10.1109/TIFS.2021.3106167}.

\bibitem[Hunker and Probst(2011)]{hunker2008insiders}
Jeffrey Hunker and Christian~W Probst.
\newblock Insiders and insider threats-an overview of definitions and mitigation techniques.
\newblock \emph{Journal of Wireless Mobile Networks, Ubiquitous Computing, and Dependable Applications}, 2\penalty0 (1):\penalty0 4--27, 2011.

\bibitem[Hurst et~al.(2024)Hurst, Lerer, Goucher, Perelman, Ramesh, Clark, Ostrow, Welihinda, Hayes, Radford, et~al.]{gpt4o}
Aaron Hurst, Adam Lerer, Adam~P Goucher, Adam Perelman, Aditya Ramesh, Aidan Clark, AJ~Ostrow, Akila Welihinda, Alan Hayes, Alec Radford, et~al.
\newblock Gpt-4o system card.
\newblock \emph{arXiv preprint arXiv:2410.21276}, 2024.

\bibitem[Hutchins et~al.(2011)Hutchins, Cloppert, and Amin]{hutchins2011intelligence}
Eric~M Hutchins, Michael~J Cloppert, and Rohan~M Amin.
\newblock Intelligence-driven computer network defense informed by analysis of adversary campaigns and intrusion kill chains.
\newblock Technical report, Lockheed Martin Corporation, 2011.

\bibitem[Inc.()]{Apple2024}
Apple Inc.
\newblock Enhanced visual search with homomorphic encryption and {PIR}.
\newblock Technical white-paper, October 2024.
\newblock Retrieved January 2025 from \url{https://www.apple.com/legal/privacy/data/en/photos/}.

\bibitem[Izacard et~al.(2021)Izacard, Caron, Hosseini, Riedel, Bojanowski, Joulin, and Grave]{contriever}
Gautier Izacard, Mathilde Caron, Lucas Hosseini, Sebastian Riedel, Piotr Bojanowski, Armand Joulin, and Edouard Grave.
\newblock Unsupervised dense information retrieval with contrastive learning.
\newblock \emph{arXiv preprint arXiv:2112.09118}, 2021.

\bibitem[Jiang et~al.(2024)Jiang, Pan, Hong, Bao, and Yang]{jiang2024ragthief}
Changyue Jiang, Xudong Pan, Geng Hong, Chenfu Bao, and Min Yang.
\newblock Rag-thief: Scalable extraction of private data from retrieval-augmented generation applications with agent-based attacks.
\newblock \emph{arXiv preprint arXiv:2411.14110}, 2024.

\bibitem[Juvekar et~al.(2018)Juvekar, Vaikuntanathan, and Chandrakasan]{Juvekar2018}
Chiraag Juvekar, Vinod Vaikuntanathan, and Abhishek Chandrakasan.
\newblock {GAZELLE}: A low latency framework for secure neural network inference.
\newblock In \emph{{USENIX} Security Symposium}, pages 1651--1669, 2018.

\bibitem[Karpukhin et~al.(2020)Karpukhin, O{\u{g}}uz, Min, Lewis, Wu, Edunov, Chen, and Yih]{dpr}
Vladimir Karpukhin, Barlas O{\u{g}}uz, Sewon Min, Patrick Lewis, Ledell Wu, Sergey Edunov, Danqi Chen, and Wen-tau Yih.
\newblock Dense passage retrieval for open-domain question answering.
\newblock \emph{arXiv preprint arXiv:2004.04906}, 2020.

\bibitem[Kojima et~al.(2022)Kojima, Gu, Reid, Matsuo, and Iwasawa]{Kojima2022}
Takeshi Kojima, Shixiang~Shane Gu, Machel Reid, Yutaka Matsuo, and Yusuke Iwasawa.
\newblock Large language models are zero-shot reasoners.
\newblock \emph{arXiv preprint arXiv:2205.11916}, 2022.
\newblock URL \url{https://arxiv.org/abs/2205.11916}.

\bibitem[Langlois and Stehl{\'e}(2015)]{mlwe}
Adeline Langlois and Damien Stehl{\'e}.
\newblock Worst-case to average-case reductions for module lattices.
\newblock \emph{Designs, Codes and Cryptography}, 75\penalty0 (3):\penalty0 565--599, 2015.

\bibitem[Lewis et~al.(2020{\natexlab{a}})Lewis, Perez, Piktus, Petroni, Karpukhin, Goyal, Küttler, Lewis, Yih, Rocktäschel, Riedel, and Kiela]{Lewis2020}
Patrick Lewis, Ethan Perez, Aleksandra Piktus, Fabio Petroni, Vladimir Karpukhin, Naman Goyal, Heinrich Küttler, Mike Lewis, Wen-tau Yih, Tim Rocktäschel, Sebastian Riedel, and Douwe Kiela.
\newblock Retrieval-augmented generation for knowledge-intensive nlp tasks.
\newblock \emph{arXiv preprint arXiv:2005.11401}, 2020{\natexlab{a}}.
\newblock URL \url{https://arxiv.org/abs/2005.11401}.

\bibitem[Lewis et~al.(2020{\natexlab{b}})Lewis, Perez, Piktus, Petroni, Karpukhin, Goyal, Küttler, Lewis, Yih, Rocktäschel, Riedel, and Kiela]{lewis2020retrieval}
Patrick Lewis, Ethan Perez, Aleksandra Piktus, Fabio Petroni, Vladimir Karpukhin, Naman Goyal, Heinrich Küttler, Mike Lewis, Wen-tau Yih, Tim Rocktäschel, Sebastian Riedel, and Douwe Kiela.
\newblock Retrieval-augmented generation for knowledge-intensive nlp tasks.
\newblock In \emph{Advances in Neural Information Processing Systems 33}, pages 9459--9474, 2020{\natexlab{b}}.

\bibitem[Li et~al.(2023)Li, Micciancio, Raykova, and Schultz-Wu]{hintlesspir}
Baiyu Li, Daniele Micciancio, Mariana Raykova, and Mark Schultz-Wu.
\newblock Hintless single-server private information retrieval.
\newblock Cryptology {ePrint} Archive, Paper 2023/1733, 2023.

\bibitem[Li et~al.(2022)Li, Shao, Wang, Guo, Xing, and Zhang]{Mao2022}
Dacheng Li, Rulin Shao, Hongyi Wang, Han Guo, Eric~P. Xing, and Hao Zhang.
\newblock {MPCFormer}: Fast, performant and private transformer inference with {MPC}.
\newblock \emph{CoRR}, abs/2211.01452, 2022.
\newblock \doi{10.48550/arXiv.2211.01452}.

\bibitem[Li et~al.(2024{\natexlab{a}})Li, Balachandran, Feng, Ilgen, Pierson, Koh, and Tsvetkov]{mediq}
Shuyue~Stella Li, Vidhisha Balachandran, Shangbin Feng, Jonathan Ilgen, Emma Pierson, Pang~Wei Koh, and Yulia Tsvetkov.
\newblock Mediq: Question-asking llms for adaptive and reliable medical reasoning.
\newblock \emph{arXiv preprint arXiv:2406.00922}, 2024{\natexlab{a}}.

\bibitem[Li et~al.(2024{\natexlab{b}})Li, Balachandran, Feng, Ilgen, Pierson, Koh, and Tsvetkov]{li2024mediq}
Shuyue~Stella Li, Vidhisha Balachandran, Shangbin Feng, Jonathan~S. Ilgen, Emma Pierson, Pang~Wei Koh, and Yulia Tsvetkov.
\newblock {MediQ}: Question-asking {LLMs} and a benchmark for reliable interactive clinical reasoning.
\newblock \emph{arXiv preprint arXiv:2406.00922}, 2024{\natexlab{b}}.

\bibitem[Lin et~al.(2023)Lin, Asai, Li, Oguz, Lin, Mehdad, Yih, and Chen]{dragon}
Sheng-Chieh Lin, Akari Asai, Minghan Li, Barlas Oguz, Jimmy Lin, Yashar Mehdad, Wen-tau Yih, and Xilun Chen.
\newblock How to train your dragon: Diverse augmentation towards generalizable dense retrieval.
\newblock \emph{arXiv preprint arXiv:2302.07452}, 2023.

\bibitem[{Lindy AI}(2024)]{LindyAI2024}
{Lindy AI}.
\newblock Lindy — meet your ai assistant.
\newblock \url{https://www.lindy.ai/lindy-agents/ai-assistant}, 2024.
\newblock Accessed 16 May 2025.

\bibitem[Liu et~al.(2023)Liu, Yu, Zhang, Xu, Lei, Lai, Gu, Ding, Men, and Yang]{liuagentbench}
Xiao Liu, Hao Yu, Hanchen Zhang, Yifan Xu, Xuanyu Lei, Hanyu Lai, Yu~Gu, Hangliang Ding, Kaiwen Men, and Kejuan Yang.
\newblock Agentbench: Evaluating llms as agents.
\newblock \emph{arXiv preprint arXiv:2308.03688}, 2023.

\bibitem[Liu et~al.(2024)Liu, Zhang, Miao, Le, and Li]{Liu2024OptLLM}
Yueyue Liu, Hongyu Zhang, Yuantian Miao, Van-Hoang Le, and Zhiqiang Li.
\newblock Optllm: Optimal assignment of queries to large language models.
\newblock \emph{arXiv preprint arXiv:2405.15130}, 2024.
\newblock URL \url{https://arxiv.org/abs/2405.15130}.

\bibitem[Liu et~al.(2025)]{liu2025edge}
Yuxuan Liu et~al.
\newblock A review on edge large language models: Design, execution, and optimization.
\newblock \emph{ACM Computing Surveys}, 1\penalty0 (1):\penalty0 1--36, 2025.

\bibitem[{LlamaIndex Team}(2024)]{LlamaIndex2024}
{LlamaIndex Team}.
\newblock Llamaindex newsletter 2024-06-11: Enhanced memory modules boost agentic rag capabilities.
\newblock \url{https://www.llamaindex.ai/blog/llamaindex-newsletter-2024-06-11}, June 2024.
\newblock Accessed 16 May 2025.

\bibitem[Maharana et~al.(2024)Maharana, Lee, Tulyakov, Bansal, Barbieri, and Fang]{locomo}
Adyasha Maharana, Dong-Ho Lee, Sergey Tulyakov, Mohit Bansal, Francesco Barbieri, and Yuwei Fang.
\newblock Evaluating very long-term conversational memory of llm agents.
\newblock \emph{arXiv preprint arXiv:2402.17753}, 2024.

\bibitem[Mavroudis(2024)]{Mavroudis2024}
Vasilios Mavroudis.
\newblock Langchain.
\newblock White paper, The Alan Turing Institute, 2024.
\newblock URL \url{https://www.turing.ac.uk/sites/default/files/2024-11/langchain.pdf}.

\bibitem[Menon and Wu(2022)]{spiralpir}
Samir~Jordan Menon and David~J. Wu.
\newblock Spiral: Fast, high-rate single-server {PIR} via {FHE} composition.
\newblock Cryptology {ePrint} Archive, Paper 2022/368, 2022.

\bibitem[Micciancio and Regev(2009)]{latticepqc}
Daniele Micciancio and Oded Regev.
\newblock Lattice-based cryptography.
\newblock In \emph{Post-quantum cryptography}, pages 147--191. Springer, 2009.

\bibitem[Mireshghallah et~al.(2023)Mireshghallah, Kim, Zhou, Tsvetkov, Sap, Shokri, and Choi]{confaide}
Niloofar Mireshghallah, Hyunwoo Kim, Xuhui Zhou, Yulia Tsvetkov, Maarten Sap, Reza Shokri, and Yejin Choi.
\newblock Can llms keep a secret? testing privacy implications of language models via contextual integrity theory.
\newblock \emph{arXiv preprint arXiv:2310.17884}, 2023.

\bibitem[Mughees et~al.(2021)Mughees, Chen, and Ren]{onionpir}
Muhammad~Haris Mughees, Hao Chen, and Ling Ren.
\newblock Onionpir: Response efficient single-server pir.
\newblock In \emph{Proceedings of the 2021 ACM SIGSAC conference on computer and communications security}, pages 2292--2306, 2021.

\bibitem[{OpenAI}(2024)]{OpenAI2024Memory}
{OpenAI}.
\newblock Memory and new controls for chatgpt.
\newblock \url{https://openai.com/index/memory-and-new-controls-for-chatgpt/}, February 2024.
\newblock Accessed 16 May 2025.

\bibitem[{OpenAI}(2025)]{OpenAI2024Operator}
{OpenAI}.
\newblock Computer-using agent: Powering operator with a universal interface.
\newblock \url{https://openai.com/index/computer-using-agent/}, January 2025.
\newblock Accessed 16 May 2025.

\bibitem[Park et~al.(2023)Park, O'Brien, Cai, Morris, Liang, and Bernstein]{Park2023}
Joon~Sung Park, Joseph~C. O'Brien, Carrie~J. Cai, Meredith~Ringel Morris, Percy Liang, and Michael~S. Bernstein.
\newblock Generative agents: Interactive simulacra of human behavior.
\newblock \emph{arXiv preprint arXiv:2304.03442}, 2023.
\newblock URL \url{https://arxiv.org/abs/2304.03442}.

\bibitem[Press et~al.(2022)Press, Zhang, Min, Schmidt, Smith, and Lewis]{Press2022}
Ofir Press, Muru Zhang, Sewon Min, Ludwig Schmidt, Noah~A. Smith, and Mike Lewis.
\newblock Measuring and narrowing the compositionality gap in language models.
\newblock \emph{arXiv preprint arXiv:2210.03350}, 2022.
\newblock URL \url{https://arxiv.org/abs/2210.03350}.

\bibitem[Qiu et~al.(2024)Qiu, Lam, Li, Acharya, Wong, Darzi, Yuan, and Topol]{qiu2024llm}
Jianing Qiu, Kyle Lam, Guohao Li, Amish Acharya, Tien~Yin Wong, Ara Darzi, Wu~Yuan, and Eric~J Topol.
\newblock Llm-based agentic systems in medicine and healthcare.
\newblock \emph{Nature Machine Intelligence}, 6\penalty0 (12):\penalty0 1418--1420, 2024.

\bibitem[Ren et~al.(2023)Ren, Chen, Gu, Lu, Zhong, Lu, Zhang, Zhang, Wu, Zheng, Liu, Chu, Hong, Wei, Niu, and Xie]{cham}
Xuanle Ren, Zhaohui Chen, Zhen Gu, Yanheng Lu, Ruiguang Zhong, Wen-Jie Lu, Jiansong Zhang, Yichi Zhang, Hanghang Wu, Xiaofu Zheng, Heng Liu, Tingqiang Chu, Cheng Hong, Changzheng Wei, Dimin Niu, and Yuan Xie.
\newblock Cham: A customized homomorphic encryption accelerator for fast matrix-vector product.
\newblock In \emph{2023 60th ACM/IEEE Design Automation Conference (DAC)}, pages 1--6, 2023.
\newblock \doi{10.1109/DAC56929.2023.10247696}.

\bibitem[{Rewind AI}(2024)]{RewindAI2024}
{Rewind AI}.
\newblock Rewind: Your ai assistant that has all the context.
\newblock \url{https://www.rewind.ai/}, 2024.
\newblock Accessed 16 May 2025.

\bibitem[Riazi et~al.(2019)Riazi, Samragh, Chen, Laine, Lauter, and Koushanfar]{Riazi2019}
M.~Sadegh Riazi, Mohammad Samragh, Hao Chen, Kim Laine, Kristin~E. Lauter, and Farinaz Koushanfar.
\newblock {XONN}: Xnor-based oblivious deep neural network inference.
\newblock In \emph{28th {USENIX} Security Symposium}, pages 1501--1518, 2019.

\bibitem[Rivest et~al.(1978)Rivest, Adleman, and Dertouzos]{rivest1978data}
Ronald~L Rivest, Len Adleman, and Michael~L Dertouzos.
\newblock On data banks and privacy homomorphisms.
\newblock \emph{Foundations of secure computation}, 4\penalty0 (11):\penalty0 169--180, 1978.

\bibitem[Rogaway(2004)]{aesind}
Phillip Rogaway.
\newblock Nonce-based symmetric encryption.
\newblock In \emph{International workshop on fast software encryption}, pages 348--358. Springer, 2004.

\bibitem[Schuhmann et~al.(2022)Schuhmann, Beaumont, Vencu, Gordon, Wightman, Cherti, Coombes, Katta, Mullis, Wortsman, et~al.]{laion}
Christoph Schuhmann, Romain Beaumont, Richard Vencu, Cade Gordon, Ross Wightman, Mehdi Cherti, Theo Coombes, Aarush Katta, Clayton Mullis, Mitchell Wortsman, et~al.
\newblock Laion-5b: An open large-scale dataset for training next generation image-text models.
\newblock \emph{Advances in Neural Information Processing Systems}, 2022.

\bibitem[Shen et~al.(2023)Shen, Song, Tan, Li, Lu, and Zhuang]{Shen2023}
Yongliang Shen, Kaitao Song, Xu~Tan, Dongsheng Li, Weiming Lu, and Yueting Zhuang.
\newblock Hugginggpt: Solving ai tasks with chatgpt and its friends in hugging face.
\newblock \emph{arXiv preprint arXiv:2303.17580}, 2023.
\newblock URL \url{https://arxiv.org/abs/2303.17580}.

\bibitem[Siyan et~al.(2024)Siyan, Raghuram, Khattab, Hirschberg, and Yu]{papillon}
Li~Siyan, Vethavikashini~Chithrra Raghuram, Omar Khattab, Julia Hirschberg, and Zhou Yu.
\newblock Papillon: Privacy preservation from internet-based and local language model ensembles.
\newblock \emph{arXiv preprint arXiv:2410.17127}, 2024.

\bibitem[Song et~al.(2025)Song, Ashktorab, and Malone]{song2025togedule}
Jaeyoon Song, Zahra Ashktorab, and Thomas~W Malone.
\newblock Togedule: Scheduling meetings with large language models and adaptive representations of group availability.
\newblock \emph{arXiv preprint arXiv:2505.01000}, 2025.

\bibitem[Staab et~al.(2024)Staab, Vero, Balunovi{\'c}, and Vechev]{staab2024large}
Robin Staab, Mark Vero, Mislav Balunovi{\'c}, and Martin Vechev.
\newblock Large language models are advanced anonymizers.
\newblock \emph{arXiv preprint arXiv:2402.13846}, 2024.

\bibitem[Su et~al.(2022)Su, Shi, Kasai, Wang, Hu, Ostendorf, Yih, Smith, Zettlemoyer, and Yu]{instructor}
Hongjin Su, Weijia Shi, Jungo Kasai, Yizhong Wang, Yushi Hu, Mari Ostendorf, Wen-tau Yih, Noah~A Smith, Luke Zettlemoyer, and Tao Yu.
\newblock One embedder, any task: Instruction-finetuned text embeddings.
\newblock \emph{arXiv preprint arXiv:2212.09741}, 2022.

\bibitem[Tang et~al.(2023)Tang, Shin, Inan, Manoel, Mireshghallah, Lin, Gopi, Kulkarni, and Sim]{tang2023privacy}
Xinyu Tang, Richard Shin, Huseyin~A Inan, Andre Manoel, Fatemehsadat Mireshghallah, Zinan Lin, Sivakanth Gopi, Janardhan Kulkarni, and Robert Sim.
\newblock Privacy-preserving in-context learning with differentially private few-shot generation.
\newblock \emph{arXiv preprint arXiv:2309.11765}, 2023.

\bibitem[Team(2025)]{fireworks}
Fireworks Team.
\newblock Fireworks api documentation, 2025.
\newblock Available at \url{https://docs.fireworks.ai/}.

\bibitem[Team et~al.(2024)Team, Georgiev, Lei, Burnell, Bai, Gulati, Tanzer, Vincent, Pan, Wang, et~al.]{gemini}
Gemini Team, Petko Georgiev, Ving~Ian Lei, Ryan Burnell, Libin Bai, Anmol Gulati, Garrett Tanzer, Damien Vincent, Zhufeng Pan, Shibo Wang, et~al.
\newblock Gemini 1.5: Unlocking multimodal understanding across millions of tokens of context.
\newblock \emph{arXiv preprint arXiv:2403.05530}, 2024.

\bibitem[Wang et~al.(2024)]{wang2024cloud}
David Wang et~al.
\newblock The pros and cons of using large language models (llms) in the cloud vs. running llms locally.
\newblock \emph{DataCamp}, 2024.

\bibitem[Wei et~al.(2022)Wei, Wang, Schuurmans, Bosma, Ichter, Xia, Chi, Le, and Zhou]{Wei2022}
Jason Wei, Xuezhi Wang, Dale Schuurmans, Maarten Bosma, Brian Ichter, Fei Xia, Ed~H. Chi, Quoc~V. Le, and Denny Zhou.
\newblock Chain-of-thought prompting elicits reasoning in large language models.
\newblock \emph{arXiv preprint arXiv:2201.11903}, 2022.
\newblock URL \url{https://arxiv.org/abs/2201.11903}.

\bibitem[Xin et~al.(2025)Xin, Mireshghallah, Li, Duan, Kim, Choi, Tsvetkov, Oh, and Koh]{xin2025false}
Rui Xin, Niloofar Mireshghallah, Shuyue~Stella Li, Michael Duan, Hyunwoo Kim, Yejin Choi, Yulia Tsvetkov, Sewoong Oh, and Pang~Wei Koh.
\newblock A false sense of privacy: Evaluating textual data sanitization beyond surface-level privacy leakage.
\newblock \emph{arXiv preprint arXiv:2504.21035}, 2025.

\bibitem[Zelikman et~al.(2022)Zelikman, Wu, Mu, and Goodman]{Zelikman2022}
Eric Zelikman, Yuhuai Wu, Jesse Mu, and Noah~D. Goodman.
\newblock Star: Bootstrapping reasoning with reasoning.
\newblock \emph{arXiv preprint arXiv:2203.14465}, 2022.
\newblock URL \url{https://arxiv.org/abs/2203.14465}.

\bibitem[Zeng et~al.(2022)Zeng, Attarian, Ichter, Choromanski, Wong, Welker, Tombari, Purohit, Ryoo, Sindhwani, Lee, Vanhoucke, and Florence]{Zeng2022}
Andy Zeng, Maria Attarian, Brian Ichter, Krzysztof Choromanski, Adrian Wong, Stefan Welker, Federico Tombari, Aveek Purohit, Michael Ryoo, Vikas Sindhwani, Johnny Lee, Vincent Vanhoucke, and Pete Florence.
\newblock Socratic models: Composing zero-shot multimodal reasoning with language.
\newblock \emph{arXiv preprint arXiv:2204.00598}, 2022.
\newblock URL \url{https://arxiv.org/abs/2204.00598}.

\bibitem[Zeng et~al.(2024{\natexlab{a}})Zeng, Zhang, He, Liu, Xing, Xu, Ren, Chang, Wang, Yin, and Tang]{zeng2024good}
Shenglai Zeng, Jiankun Zhang, Pengfei He, Yiding Liu, Yue Xing, Han Xu, Jie Ren, Yi~Chang, Shuaiqiang Wang, Dawei Yin, and Jiliang Tang.
\newblock The good and the bad: Exploring privacy issues in retrieval-augmented generation ({RAG}).
\newblock In \emph{Findings of the Association for Computational Linguistics: ACL 2024}, pages 4505--4524, 2024{\natexlab{a}}.

\bibitem[Zeng et~al.(2024{\natexlab{b}})Zeng, Zhang, He, Xing, Liu, Xu, Ren, Wang, Yin, Chang, et~al.]{zeng2024privacy}
Shenglai Zeng, Jiankun Zhang, Pengfei He, Yue Xing, Yiding Liu, Han Xu, Jie Ren, Shuaiqiang Wang, Dawei Yin, Yi~Chang, et~al.
\newblock The good and the bad: Exploring privacy issues in retrieval-augmented generation (rag).
\newblock \emph{arXiv preprint arXiv:2402.16893}, 2024{\natexlab{b}}.

\bibitem[Zheng et~al.(2024)Zheng, Chen, Han, and Zheng]{Zhu2024}
Fei Zheng, Chaochao Chen, Zhongxuan Han, and Xiaolin Zheng.
\newblock {PermLLM}: Private inference of large language models within 3 seconds under {WAN}.
\newblock \emph{CoRR}, abs/2405.18744, 2024.
\newblock \doi{10.48550/arXiv.2405.18744}.

\bibitem[Zhou et~al.(2022)Zhou, Sch{\"a}rli, Hou, Wei, Scales, Wang, Schuurmans, Cui, Bousquet, Le, and Chi]{Zhou2022}
Denny Zhou, Nathanael Sch{\"a}rli, Le~Hou, Jason Wei, Nathan Scales, Xuezhi Wang, Dale Schuurmans, Claire Cui, Olivier Bousquet, Quoc~V. Le, and Ed~H. Chi.
\newblock Least-to-most prompting enables complex reasoning in large language models.
\newblock \emph{arXiv preprint arXiv:2205.10625}, 2022.
\newblock URL \url{https://arxiv.org/abs/2205.10625}.

\bibitem[Zhu et~al.(2024)Zhu, Patel, Zaharia, and Popa]{compass}
Jinhao Zhu, Liana Patel, Matei Zaharia, and Raluca~Ada Popa.
\newblock Compass: Encrypted semantic search with high accuracy.
\newblock Cryptology {ePrint} Archive, Paper 2024/1255, 2024.

\end{thebibliography}
